\documentclass[twocolumn,prd,superscriptaddress,preprintnumbers,nofootinbib]{revtex4-2}

\usepackage{comment} 

\usepackage{graphicx}
\usepackage{amsmath,bm,amssymb,amsfonts,dsfont}
\usepackage[usenames,dvipsnames]{xcolor}
\usepackage[normalem]{ulem}
\usepackage{url}
\usepackage{array}
\usepackage{booktabs}
\usepackage{multirow}
\usepackage{float}
\usepackage[colorlinks  = true,
            linkcolor   = NavyBlue,
            urlcolor    = NavyBlue,
            citecolor   = NavyBlue,
            anchorcolor = NavyBlue]{hyperref}
\usepackage{cprotect}

\usepackage{amsmath}

\usepackage{appendix}

\usepackage[switch]{lineno}

\usepackage{tikz-feynman}
\tikzfeynmanset{compat=1.1.0}

\renewcommand{\arraystretch}{1.5}

\pdfoutput=1 

\newcommand{\lsim}{\mathrel{\mathop{\kern 0pt \rlap
  {\raise.2ex\hbox{$<$}}}
  \lower.9ex\hbox{\kern-.190em $\sim$}}}
\newcommand{\gsim}{\mathrel{\mathop{\kern 0pt \rlap
  {\raise.2ex\hbox{$>$}}}
  \lower.9ex\hbox{\kern-.190em $\sim$}}}

\newcommand{\pp}{pp}

\newcommand{\dbar}{\ensuremath{\overline{\mathrm{D}}}}

\newcommand{\hebar}{$^3\overline{\mathrm{He}}$}

\tikzfeynmanset{warn luatex=false} 

\begin{document}



\title{Toward universal coalescence models for antideuteron production}

\author{Mattia Di Mauro}
\email{dimauro.mattia@gmail.com}
\affiliation{Istituto Nazionale di Fisica Nucleare, Sezione di Torino, Via P. Giuria 1, 10125 Torino, Italy}

\author{Jordan Koechler}
\email{jordan.koechler@gmail.com}
\affiliation{Istituto Nazionale di Fisica Nucleare, Sezione di Torino, Via P. Giuria 1, 10125 Torino, Italy}

\author{Lorenzo Stefanuto}
\email{lorenzo.stefanuto@unito.it}
\affiliation{Department of Physics, University of Torino, Via P. Giuria, 1, 10125 Torino, Italy}
\affiliation{Istituto Nazionale di Fisica Nucleare, Sezione di Torino, Via P. Giuria 1, 10125 Torino, Italy}

\author{Francesca Bellini}
\email{f.bellini@unibo.it}
\affiliation{Dipartimento di Fisica e Astronomia, Universit\`a di Bologna, via Irnerio 46, 40126 Bologna, Italy}
\affiliation{Istituto Nazionale di Fisica Nucleare, Sezione di Bologna, viale Berti Pichat 6/2, 40127, Bologna, Italy}

\author{Fiorenza Donato}
\email{fiorenza.donato@unito.com}
\affiliation{Department of Physics, University of Torino, Via P. Giuria, 1, 10125 Torino, Italy}
\affiliation{Istituto Nazionale di Fisica Nucleare, Sezione di Torino, Via P. Giuria 1, 10125 Torino, Italy}

\author{Nicolao Fornengo}
\email{nicolao.fornengo@unito.com}
\affiliation{Department of Physics, University of Torino, Via P. Giuria, 1, 10125 Torino, Italy}
\affiliation{Istituto Nazionale di Fisica Nucleare, Sezione di Torino, Via P. Giuria 1, 10125 Torino, Italy}

\begin{abstract}
Cosmic-ray (CR) antinuclei, especially antideuteron $\overline{\rm D}$ and antihelium-3 nuclei ${}^3\overline{\rm He}$, are among the most promising messengers for indirect dark matter (DM) searches. This is because secondary production in CR interactions with the interstellar medium is strongly suppressed at kinetic energies $K\simeq (0.1 - 1)$~GeV/$n$, typically one to two orders of magnitude below fluxes expected in standard DM scenarios. 
From the theoretical side, the formation of $\overline{\rm D}$ and ${}^3\overline{\rm He}$ is governed by coalescence, whose dynamics cannot yet be reliably derived from first principles. Phenomenological approaches therefore introduce effective coalescence parameters, possibly dependent on collision energy and production environment (hadronic versus electroweak). In this work we show that a common set of physically motivated coalescence models can simultaneously reproduce collider data in two qualitatively different regimes: \textsf{ALICE} measurements of (anti)deuteron production in $pp$ collisions at $\sqrt{s}=(0.9 - 13)$~TeV and the \textsf{ALEPH} $\overline{\rm D}$ multiplicity in hadronic $Z$ decays at $\sqrt{s}=m_Z$. 
We test both simple event-by-event prescriptions based on a relative-momentum cutoff, finding a preferred coalescence scale $p_{\rm coal}\simeq 0.2$~GeV, and quantum-mechanical models in the Wigner formalism. In the latter, a Gaussian bound-state wavefunction gives a best-fit momentum width, corresponding to $\delta\simeq 1.7$~fm, while a parameter-free implementation using the Argonne $v_{18}$ wavefunction (constrained by proton-neutron scattering data) agrees with \textsf{ALICE} spectra at the $\sim 25\%$ level. Overall, our results support an approximately universal coalescence description across energies and production environments, strengthening the theoretical basis for interpreting upcoming CR antinuclei searches.
\end{abstract}


\maketitle
\newpage

\flushbottom

\section{Introduction}

Cosmic-ray (CR) light antinuclei, especially antideuterons (\dbar) and antihelium nuclei (\hebar), are considered among the cleanest messengers for indirect dark matter (DM) searches. The key reason is the pronounced suppression, at kinetic energy per nucleon $K\lsim 1~\mathrm{GeV}/n$, of the secondary production associated with interactions of primary CRs with the interstellar medium (ISM). In fact, baryon-number conservation implies a large kinematic threshold for producing $\dbar$ in $\pp$ collisions: the projectile kinetic energy must satisfy $K\gtrsim 16\,m_p$, such that secondary $\dbar$ fluxes are strongly depleted below $\sim 1~\mathrm{GeV}/n$. 
Instead, DM annihilation or decay in the Galactic halo can populate the sub-GeV$/n$ regime more efficiently, since they are not subject to the same threshold suppression. Theoretical predictions for the \dbar\ flux from standard weakly interacting massive particles (WIMPs) can typically be up to two orders of magnitude higher than the secondary production expectations at low $K$ \cite{Donato:1999gy,Fornengo:2013osa,Korsmeier:2017xzj,Kachelriess:2020uoh,Serksnyte:2022onw,DeLaTorreLuque:2024htu}. Consequently, even a small number of detected events at these energies would be difficult to reconcile with standard astrophysical backgrounds and could provide a distinctive DM signature.

Despite sustained experimental efforts over the past decades, no conclusive detection of cosmic antinuclei has been established \cite{Abe:2012tz,PAMELA:2017bna,AMS:2021nhj,Osteria:2020lxn,vonDoetinchem:2020vbj,PhysRevLett.132.131001}. The strongest published constraint on the antideuteron flux is currently provided by the \textsf{BESS-Polar II} experiment \cite{PhysRevLett.132.131001}, which sets an upper limit of $6.7\times10^{-5}\,\mathrm{(m^2\,s\,sr\,GeV}/n)^{-1}$ in the range $K\in(0.163,\,1.100)\,\mathrm{GeV}/n$. In the near future, the situation is expected to improve substantially: \textsf{AMS-02} on the International Space Station \cite{2008ICRC....4..765C} has been taking data since 2011 and is expected to reach, by about 2030, a sensitivity of $(2$--$8)\times10^{-7}\,\mathrm{(m^{2}\,s\,sr\,GeV}/n)^{-1}$ in the interval $K\in(0.2,\,4)~\mathrm{GeV}/n$. Moreover, the balloon-borne \textsf{GAPS} program \cite{Aramaki:2015laa,Osteria:2020lxn,vonDoetinchem:2020vbj} is designed to reach peak sensitivity comparable to \textsf{AMS-02} precisely in the low-energy window $K\in(0.1,\,1)~\mathrm{GeV}/n$ where the secondary background is minimal \cite{vonDoetinchem:2015yva}, and has just completed its first flight (25 days) \cite{gaps_Hailey_cern}. These searches therefore have the potential to access flux levels where DM signals can emerge over astrophysical production.

Even though antinuclei may provide one of the cleanest channels for DM indirect detection among cosmic messengers, the theoretical description of their production is affected by significant uncertainties. Interpreting current limits and translating future sensitivities into robust constraints therefore requires accurate predictions for the antinucleus source term. A widely used framework is provided by \emph{coalescence models} \cite{Kapusta:1980,Butler:1963,Scheibl:1998tk}, in which antinuclei form when the constituent antinucleons (antiprotons and antineutrons) are sufficiently close in phase space to fuse into a bound state.

One of the most relevant implications of coalescence, together with Standard Model hadronization, is that the antinuclei formation probability is strongly suppressed with increasing mass number. For instance, in a representative Monte Carlo setup, the relative abundances from DM annihilation scale steeply from antiprotons to $\dbar$ and to ${}^3\overline{\mathrm{He}}$ \cite{DiMauro:2025vxp}:
\begin{equation}
\label{eq:ratios}
\overline{p} : \overline{\mathrm{D}} : {}^3\overline{\mathrm{He}} \;\sim\; 1 : 1.4\times10^{-4} : 3.4\times10^{-8}~.
\end{equation}
This means that the formation probability drops drastically, by about a factor of $10^4$, for each additional antinucleon in the nucleus. Standard WIMP models that would be able to produce observable fluxes of $\overline{\rm D}$ and ${}^3\overline{\rm He}$ would also predict antiproton fluxes from DM that should already have been detected by \textsf{AMS-02}, on top of the known secondary astrophysical component~\cite{Korsmeier:2017xzj}.

In the simplest implementation---often referred to as the \emph{spherical coalescence} approach---a bound state is produced if, in the pair center-of-mass frame, the relative momentum satisfies $\Delta p < p_{\rm coal}$, where $p_{\rm coal}$ is the coalescence momentum. This parameter cannot be computed reliably from first principles and is instead calibrated on accelerator measurements when available \cite{Fornengo:2013osa,Kachelriess:2020uoh}. In phenomenological applications it is also common to allow for different effective values of $p_{\rm coal}$ depending on the production environment, e.g.\ DM annihilation/decay versus secondary production in hadronic CR interactions. For example, Ref.~\cite{Korsmeier:2017xzj} adopts $p_{\rm coal}\sim (200\text{--}260)~\mathrm{MeV}$ for secondary production and $p_{\rm coal}\sim 160~\mathrm{MeV}$ for DM annihilation (consistent with the calibration in Ref.~\cite{Fornengo:2013osa}).

More refined coalescence models, based on a full quantum-mechanical treatment with Wigner formalism, account for the dependence on the momentum distribution of the nucleons, the nucleus wave function, and the characteristics of the nucleon-emitting source~\cite{Bellini:2018epz,Blum:2017qnn,Bellini:2020cbj,Kachelriess:2019taq,Kachelriess:2020uoh}. This approach is the foundation of recent developments that calculate the production of (anti)nuclei in hadronic interactions event-by-event by employing Monte Carlo simulations~\cite{Kachelriess:2019taq,Kachelriess:2023jis,Horst:2023oti}.

In this context, collider measurements provide crucial external anchors to tune the coalescence model. The \textsf{ALICE} collaboration has measured light (anti)nuclei production in $\pp$ collisions over a wide range of center-of-mass energies and multiplicities, delivering multi-differential information in transverse momentum and event activity \cite{ALICE:2015wav,ALICE:2017xrp,ALICE:2019dgz,ALICE:2020foi,ALICE:2021mfm,ALICE:2021ovi}. On the other hand, \textsf{ALEPH} has measured the $\overline{\mathrm D}$ multiplicity in hadronic $Z$ decays \cite{2006192,ALEPH:2006qoi}, a final state often taken as a close proxy for antinucleon production from DM annihilation into quarks. These datasets probe distinct underlying environments---hadronic $\pp$ interactions versus hadronic fragmentation in $e^+e^-$ collisions---and are therefore ideally suited to stress-test the robustness of coalescence modeling.

Recently, in Ref.~\cite{DiMauro:2024kml} some of the authors used the \textsf{ALEPH} $\overline{\mathrm D}$ measurement \cite{2006192,ALEPH:2006qoi} to calibrate several coalescence prescriptions and, in turn, to significantly reduce the systematic uncertainty affecting the predicted DM-induced $\overline{\mathrm D}$ source spectra. In particular, they employed a model that implements the Argonne nucleon--nucleon potential, with explicit charge dependence and charge asymmetry \cite{Wiringa:1994wb}, in a Wigner formalism (hereafter {\tt Argonne}) for a full quantum-mechanical treatment of the coalescence process. The Argonne potential is tuned directly to $p$--$p$ and $n$--$p$ inelastic scattering data, to low-energy $n$--$n$ scattering, and to the deuteron binding energy \cite{Wiringa:1994wb}. This approach is more predictive than purely phenomenological prescriptions since the nucleon--nucleon potential is fixed by scattering data and, in this implementation, the coalescence mechanism does not introduce additional free parameters, while providing excellent agreement with the \textsf{ALEPH} measurement \cite{ALEPH:2006qoi} of the $\overline{\rm D}$ multiplicity. They also show that, once the antinucleon yield is properly calibrated and the coalescence model is tuned on a common dataset, the residual theoretical systematics become negligible compared to other uncertainties, such as those from QCD modeling, the DM density distribution, or CR propagation.

Building on this progress, the purpose of this work is twofold. On the one hand, we apply the analysis method of Ref.~\cite{DiMauro:2024kml} to the secondary production of antinuclei. On the other hand, we directly test the \emph{universality} of the coalescence mechanism across production processes and energies. Concretely, we investigate whether a single, common coalescence framework can simultaneously reproduce (i) the light (anti)nuclei measurements in $\pp$ collisions by \textsf{ALICE} at TeV center-of-mass energies \cite{ALICE:2015wav,ALICE:2017xrp,ALICE:2019dgz,ALICE:2020foi,ALICE:2021mfm,ALICE:2021ovi} and (ii) the $\overline{\mathrm D}$ multiplicity in hadronic $Z$ decays measured by \textsf{ALEPH} \cite{2006192,ALEPH:2006qoi}. Establishing (or falsifying) such universality is critical for consolidating collider-calibrated antinucleus production models into a coherent input for cosmic-ray antinuclei searches with \textsf{AMS-02} and \textsf{GAPS} \cite{2008ICRC....4..765C,Aramaki:2015laa,vonDoetinchem:2015yva}.

The paper is organized as follows: In Sec.~\ref{sec:model} we describe the coalescence model we employ. Section~\ref{sec:tuning} describes the tuning of the \texttt{PYTHIA} Monte Carlo code that we use to calibrate the production of antinucleons. Section~\ref{sec:pythiaDbar} contains the description of the assumptions adopted for the antiproton/antineutron production vertices, and we investigate their production from resonance-baryon decays and from prompt hadronization. In Sec.~\ref{sec:results} we present the results, in Sec.~\ref{sec:source_secondary} we compute the source spectrum in the Galaxy, and we conclude in Sec.~\ref{sec:conclusions}.

\section{Coalescence models}
\label{sec:model}

Light (anti)nuclei formation is commonly described in terms of \emph{coalescence}: a bound state is produced if the antinucleons are sufficiently close in phase space. In practice, this idea is implemented at different levels of sophistication, ranging from analytic ``spherical'' prescriptions to event-by-event Monte-Carlo afterburners, up to fully quantum-mechanical formulations based on Wigner functions. In the following we summarize the approaches relevant for this work, emphasizing their assumptions and domains of applicability.

\subsection{Spherical approach}
\label{subsec:spherical}

The most elementary coalescence implementation assumes that (anti)protons and (anti)neutrons are emitted independently and isotropically, and that correlations between them can be neglected \cite{Donato:1999gy,Korsmeier:2017xzj}. Under these hypotheses, the formation probability depends on a single parameter, the \emph{coalescence momentum} $p_{\rm coal}$, and the (anti)nucleus spectrum can be related analytically to the single-(anti)nucleon spectra. For a nucleus with mass number $A$ and proton/neutron numbers $(Z,N)$, one writes
\begin{equation}
E_A \frac{d^3 N_A}{dp_A^3}
= B_A
\left( E_p \frac{d^3 N_p}{dp_p^3} \right)^Z
\left( E_n \frac{d^3 N_n}{dp_n^3} \right)^N ,
\end{equation}
where the nucleon momenta are evaluated at $p_{p,n}=p_A/A$ and $E_{p,n}$ denote the corresponding energies. In practice, the neutron spectrum is often taken to be equal to the proton one (and analogously for antiparticles), so that for antideuterons one obtains
\begin{equation}
E_{\overline{\rm D}} \frac{d^3 N_{\overline{\rm D}}}{dp_{\overline{\rm D}}^3}
= B_2 \left( E_{\bar{p}} \frac{d^3 N_{\bar{p}}}{dp_{\bar{p}}^3} \right)^2 ,
\end{equation}
with $B_2 \equiv B_{A=2}$.

In the spherical coalescence approximation, $B_A$ can be expressed in terms of the coalescence momentum $p_{\rm coal}$ as \cite{Donato:1999gy,Korsmeier:2017xzj}
\begin{equation}
\label{eq:BA_spherical}
B_A \;=\;
S_A\,
\frac{m_A}{m_p^{Z}\,m_n^{N}}\,
\left(\frac{4\pi}{3}\,p_{\rm coal}^{\,3}\right)^{A-1},
\end{equation}
where $m_A$ is the nucleus mass and $m_{p,n}$ are the proton and neutron masses. The factor $S_A$ accounts for spin/isospin combinatorics (for $\overline{\rm D}$ one typically has $S_2=3/8$).%
\footnote{Different conventions for $p_{\rm coal}$ (e.g.\ using the relative momentum $\vec q=(\vec p_p-\vec p_n)/2$) change only the overall numerical prefactor in Eq.~\eqref{eq:BA_spherical}.}
Owing to its simplicity, this prescription has been widely adopted in astroparticle applications, since once $p_{\rm coal}$ is fixed one can compute $\rm{D}$ or $\dbar$ spectra analytically from the (anti)nucleon spectra.

From the experimental side, the coalescence parameter is extracted to be of order $B_2 \sim \mathcal{O}(10^{-2})~\mathrm{GeV}^2/c^3$ in $\pp$ collisions at \textsf{LHC} energies in the transverse momentum range $p_T/A \sim 1~\mathrm{GeV}/c$, and it exhibits a rise with $p_T/A$ in multiplicity-integrated samples \cite{ALICE:2020foi,ALICE:2025antideuteronRapidity13TeV}. 
In larger collision systems and/or at lower beam energies (e.g.\ heavy-ion data at \textsf{LHC}/\textsf{RHIC}/\textsf{SPS}), smaller values are commonly found, consistent with the interpretation of $B_A$ as being inversely related to an effective emission volume \cite{ALICE:2022veq, STAR:2019sjh,NA49:2012dbarPbPb158A}.
A residual $p_T$ dependence of $B_2$ can also arise from spectral-shape effects: if the (anti)proton spectrum is harder than the (anti)deuteron one (i.e.\ the shapes are not self-similar), then a constant-$B_2$ ansatz cannot reproduce both slopes simultaneously and the extracted $B_2(p_T)$ acquires an effective $p_T$ dependence \cite{ALICE:2019multiplicitypPbLightNuclei,ALICE:2020foi}.

Despite its simplicity, neglecting correlations in momentum and configuration space can lead to biased predictions in realistic hadronic environments. In particular, the spherical approximation is known to fail in regions where correlations are most important (e.g.\ at high-$p_T$) \cite{Fornengo:2013osa,DiMauro:2024kml}. These limitations motivate the use of event-by-event Monte-Carlo implementations discussed in the next sections.

\subsection{Monte-Carlo coalescence with phase-space cutoffs}
\label{subsec:cutoff_coal}

Monte-Carlo afterburners provide an event-by-event realization of coalescence and, by construction, avoid assuming isotropy or uncorrelated particle production \cite{Fornengo:2013osa,DiMauro:2024kml}. In their simplest form, they implement coalescence through geometric cuts in phase space.

A standard choice is to require that, in the pair center-of-mass frame, the relative momentum of a proton--neutron pair lies within a sphere of radius $p_{\rm coal}$,
\begin{equation}
|\vec{p}_p-\vec{p}_n| < p_{\rm coal}\,.
\end{equation}
When applied to each generated event, this criterion selects candidate pairs solely based on their kinematics. For this reason we refer to it as the $\Delta p$ coalescence model. Owing to its minimal assumptions and computational simplicity, the $\Delta p$ prescription has long been a default tool for modeling light antinuclei formation in high-energy processes \cite{Ibarra:2012cc,Fornengo:2013osa}.

A straightforward extension incorporates spatial correlations by supplementing the momentum cut with a coordinate-space requirement,
\begin{equation}
|\vec{r}_p-\vec{r}_n| < r_{\rm coal}\,,
\end{equation}
where $r_{\rm coal}$ parametrizes the effective spatial proximity needed for binding. The scale of $r_{\rm coal}$ is expected to be of the order of the physical size of the $\overline{\rm D}$ nucleus \cite{CREMA:2016idx}, i.e.\ $r_{\rm coal}\sim 2$--$3~\mathrm{fm}$. We denote this combined criterion as the $\Delta p+\Delta r$ coalescence model. While still phenomenological, it accounts at least qualitatively for the fact that coalescence is sensitive to both relative momentum and emission geometry. In particular, it naturally suppresses the coalescence of antinucleons originating from displaced vertices (e.g.\ from weakly decaying hadrons) with promptly produced antinucleons, since such pairs typically fail the spatial cut \cite{Fornengo:2013osa,DiMauro:2025vxp}.

The $\Delta p$ and $\Delta p+\Delta r$ prescriptions are historically the most widely used event-by-event coalescence models and were introduced to overcome the main limitations of the spherical approach by retaining kinematic (and, in the second case, geometric) correlations from the underlying event generator \cite{Ibarra:2012cc,Fornengo:2013osa}.

\subsection{Monte-Carlo coalescence in the Wigner formalism}
\label{subsec:wigner_model}

To address the shortcomings of purely geometric prescriptions---notably the lack of a first-principles determination of $p_{\rm coal}$ and $r_{\rm coal}$, their possible dependence on process and energy, and the incomplete treatment of the space--time structure of the emitting source---one can formulate coalescence within a quantum-mechanical framework based on Wigner functions \cite{Scheibl:1998tk,Bellini:2018epz,Blum:2017qnn,Bellini:2020cbj,Kachelriess:2019taq,Mahlein:2023fmx}. In this approach, antinucleons are described by quantum states with non-trivial distributions in both coordinate and momentum space, and the formation probability of a bound state is obtained by projecting the two-particle state onto the deuteron bound-state density matrix.

The (anti)deuteron differential spectrum can be written as
\begin{equation}
\frac{d^3 N_{\rm D}}{d p_{\rm D}^3}
=\mathrm{Tr}\!\left(\rho_{\rm D}\,\rho_{pn}\right),
\end{equation}
with $\rho_{\rm D}=|\phi_{\rm D}\rangle\langle\phi_{\rm D}|$ and $\rho_{pn}=|\psi_{pn}\rangle\langle\psi_{pn}|$, where $\phi_{\rm D}$ is the (anti)deuteron wavefunction and $\psi_{pn}$ the two-particle $p$--$n$ wavefunction.

Introducing the relative and center-of-mass coordinates,
\begin{equation}
\vec r \equiv \vec r_{p}-\vec r_{n},\qquad
\vec R \equiv \frac{\vec r_{p}+\vec r_{n}}{2}\equiv \vec r_{\rm D},
\end{equation}
one may factorize the bound-state wavefunction into internal and center-of-mass parts as
\begin{equation}
\phi_{\rm D}(\vec r,\vec R;\vec p_{\rm D}) \propto
\varphi_{\rm D}(\vec r)\,e^{i \vec p_{\rm D}\cdot \vec R},
\end{equation}
where $\varphi_{\rm D}(\vec r)$ denotes the \emph{internal} wavefunction (depending on the relative coordinate $\vec r$), and $\vec p_{\rm D}$ is the deuteron three-momentum. The differential spectrum can then be expressed as \cite{Scheibl:1998tk,Kachelriess:2019taq}
\begin{multline}
\label{eq:wignerDspectrum_rephr}
\frac{d^3 N_{\rm D}}{d p_{\rm D}^3} =
S \int \frac{d^3 r_{\rm D}\, d^3 r\, d^3 q}{(2 \pi)^6}\;
\mathcal{D}(\vec{r}, \vec{q}) \times \\
\times W_{pn}\!\left(\frac{\vec{p}_{\rm D}}{2}+\vec{q}, \frac{\vec{p}_{\rm D}}{2}-\vec{q}, \vec{r}_{p}, \vec{r}_{n}\right),
\end{multline}
where $S$ accounts for spin/isospin statistics (for $\rm D$, $S=3/8$), $\vec{q}\equiv(\vec{p}_{p}-\vec{p}_{n})/2$ (so that $\Delta\vec p=2\vec q$), and $\vec r\equiv \vec r_{p}-\vec r_{n}$ as above. 
In the rest of the paper we will use both $\Delta r$ and $r$ to define the modulus of the vector $\vec r\equiv \vec r_{p}-\vec r_{n}$, i.e.~the distance between $p$ and $n$.
The function $W_{pn}$ is the Wigner function of the $p$--$n$ pair, while $\mathcal{D}$ is the deuteron Wigner function, defined by \cite{Scheibl:1998tk}
\begin{equation}
\mathcal{D}(\vec{r}, \vec{q})=\int d^3 \xi\;
e^{-i \vec{q} \cdot \vec{\xi}}\;
\varphi_{\rm D}(\vec{r}+\vec{\xi} / 2)\,
\varphi_{\rm D}^*(\vec{r}-\vec{\xi} / 2),
\end{equation}
and normalized as
\begin{equation}
\label{eq:Wignormalized_rephr}
\int d^3r \int \frac{d^3q}{(2 \pi)^3}\,\mathcal{D}(\vec{r}, \vec{q}) = 1 .
\end{equation}
The functional form of $\mathcal{D}(\vec r,\vec q)$ is set by the choice of $\varphi_{\rm D}$; below we consider two standard options.

For the Wigner function of the $p$--$n$ pair we adopt the factorized form
\begin{equation}
W_{pn}=
H_{pn}\!\left(\vec{r}_{p}, \vec{r}_{n}\right)\,
G_{pn}\!\left(\frac{\vec{p}_{\rm D}}{2}+\vec{q}, \frac{\vec{p}_{\rm D}}{2}-\vec{q}\right),
\end{equation}
where $G_{\rm pn}$ is the two-particle momentum distribution obtained from the Monte-Carlo generator (including correlations). For the spatial part, we approximate
\begin{equation}
H_{pn}\!\left(\vec{r}_{p}, \vec{r}_{n}\right)
= h(\vec r_{p})\,h(\vec r_{n}) ,
\end{equation}
with $h$ extracted from the same generator. We stress that this approximation neglects genuine two-particle spatial correlations beyond those implicitly induced through the common event geometry. The corresponding impact is explored through variations of the vertex modeling discussed in Sec.~\ref{sec:pythiaDbar}. With these ingredients specified, the remaining model dependence resides primarily in the bound-state wavefunction and in the effective source size.

\subsubsection{Gaussian wavefunction}
\label{sec:gaussian}

As a first benchmark (labeled as {\tt Gaussian}), we adopt a Gaussian ansatz for the internal deuteron wavefunction,
\begin{equation}
\varphi_{\rm D}(r)=\left(\pi d^2\right)^{-3/4}\exp\!\left(-\frac{r^2}{2d^2}\right),
\label{eq:deuteron_wf_Gaus}
\end{equation}
where $r\equiv|\vec r|$ denotes the modulus of the relative separation $\vec r=\vec r_{p}-\vec r_{n}$. The size parameter $d$ can be related to the RMS charge radius (defined through $r_{\rm RMS}^2=\int d^3r\,(r/2)^2|\varphi_{\rm D}(r)|^2$); in particular, $d\simeq 3.2~\mathrm{fm}$ reproduces the measured deuteron RMS charge radius \cite{CREMA:2016idx}. We further assume Gaussian single-particle source profiles,
\begin{equation}
h(r_{p,n})=\frac{1}{(2\pi\sigma^2)^{3/2}}\exp\!\left(-\frac{r_{p,n}^2}{2\sigma^2}\right),
\label{eq:source_gaus}
\end{equation}
where $\sigma$ characterizes the spatial extent of the emitting region. For a Gaussian bound state, the corresponding Wigner function takes the simple form \cite{Kachelriess:2020uoh}
\begin{equation}
\mathcal{D}(r,q)=8\,\exp\!\left(-\frac{r^2}{d^2}\right)\exp\!\left(-q^2 d^2\right).
\label{eq:deutwignerinternal}
\end{equation}

Following the common procedure in \cite{Kachelriess:2019taq,Horst:2023oti}, one may integrate out the explicit coordinate dependence in Eq.~\eqref{eq:wignerDspectrum_rephr}. Under the Gaussian assumptions above, this yields
\begin{equation}
\frac{d^{3}N_{\rm D}}{dp_{\rm D}^3}
=
\frac{3}{(2\pi)^6}
\left(\frac{d^2}{d^2+4\sigma^2}\right)^{3/2}
\int d^{3}q\;
e^{-q^2 d^2}\;
G_{pn}(\vec{p}_{\rm D},\vec{q}) ,
\label{eq:singlegaussian}
\end{equation}
so that the spectrum is obtained numerically from the Monte-Carlo estimate of $G_{pn}$, while the prefactor depends only on the deuteron size $d$ and the source size $\sigma$.

This ``integrated'' implementation is the one considered for example in \cite{Kachelriess:2019taq,Kachelriess:2020uoh,Kachelriess:2023jis}, which takes from the Monte Carlo only the momentum distribution of the antinucleons.
Instead, our goal is to exploit the full $\bar{p}$--$\bar{n}$ phase space information available in the event generator including their spatial distribution. To this end, we interpret the Wigner kernel as a probability density in $(\Delta\vec r,\Delta\vec p)$ and adopt a Gaussian probability distribution function (PDF),
\begin{equation}
\mathrm{PDF}(r,q)=N\,
\exp\!\left(-\frac{r^2}{2\sigma^2}\right)\,
\exp\!\left(-\frac{q^2\delta^2}{2}\right),
\label{eq:singlegaussianPDF}
\end{equation}
with $N$ fixed by normalization. In the strict Gaussian Wigner limit one expects $\sigma$ and $\delta$ to be respectively of order $d\sqrt{2}\simeq 4~\mathrm{fm}$ and $d/\sqrt{2}\simeq 2~\mathrm{fm}$ (cf.\ Eqs.~\eqref{eq:deutwignerinternal} and \eqref{eq:singlegaussianPDF}). However Eq.~\eqref{eq:singlegaussianPDF} should be interpreted as a generalization of Eq.~\eqref{eq:deutwignerinternal}, thus we first assume that $\sigma$ and $\delta$ are uncorrelated. The cumulative coalescence probability for a pair with separation $r$ and relative momentum $q$ can then be written as
\begin{equation}
\label{eq:coalProb}
\mathcal{P}(r,q)=\int_{0}^{r}\!\!dr'\int_{0}^{q}\!\!dq'\;\mathcal{D}(r',q'),
\end{equation}
which is consistent with the normalization in Eq.~\eqref{eq:Wignormalized_rephr}.

\subsubsection{Argonne wavefunction}
\label{sec:argonne}

As a second (labeled as {\tt Argonne}), more realistic choice for the bound-state dynamics, we consider the deuteron wavefunction associated with the Argonne $v_{18}$ potential, which is constrained by nucleon--nucleon scattering data and the deuteron binding energy \cite{Wiringa:1994wb}. In this case one may write
\begin{equation}
\varphi_{\rm D}(\vec{r})=
\frac{1}{\sqrt{4 \pi}\, r}\left[
u(r)+\frac{1}{\sqrt{8}}\, w(r)\, S_{12}(\hat{r})
\right]\chi_{1 m},
\end{equation}
where $S_{12}(\hat{r})=3(\vec{\sigma}_1\!\cdot\!\hat r)(\vec{\sigma}_2\!\cdot\!\hat r)-(\vec{\sigma}_1\!\cdot\!\vec{\sigma}_2)$ is the tensor operator, $\sigma_i$ are Pauli matrices, $\chi_{1m}$ is the spinor, and $u(r)$ and $w(r)$ are the radial $\mathrm S$- and $\mathrm D$-wave components. The normalization condition reads
\begin{equation}
\int d^3r\,|\varphi_{\rm D}(\vec r)|^2
=
\int dr\,\left[u^2(r)+w^2(r)\right]=1 .
\end{equation}
For this wavefunction the Wigner kernel $\mathcal{D}(r,q)$ does not admit a simple analytic expression; in our implementation we use a tabulated representation of $\mathcal{D}(r,q)$ as a function of $(r,q)$ based on the fit provided in \cite{Horst:2023oti}.

In our event-by-event implementation, we interpret $\mathcal{D}(\Delta r,\Delta p)$ as an acceptance weight in $(\Delta r,\Delta p)$ space. For every $\bar p$--$\bar n$ pair we evaluate $\mathcal{D}(\Delta r,\Delta p)$, draw a random number $u\in[0,1]$, and form a $\overline{\rm D}$ if $u$ is smaller than the (properly normalized) weight. This is the same method employed in Ref.~\cite{DiMauro:2024kml}.

\section{PYTHIA tuning for (anti)nucleon production}
\label{sec:tuning}

A reliable prediction of light (anti)nuclei yields from secondary production requires, as a first step, an accurate Monte-Carlo description of the underlying (anti)nucleon production in $pp$ collisions. This is particularly important in our setup, where (anti)nuclei are formed by applying the coalescence prescription directly to the event record. Throughout this work we use \texttt{PYTHIA}~8.315 \cite{Bierlich:2022pfr} as our baseline generator and implement the coalescence procedure at the Monte-Carlo level. In Ref.~\cite{DiMauro:2024kml} we followed an analogous strategy for DM annihilation, by first tuning \texttt{PYTHIA} to hadron production data at the \textsf{LEP} in $e^+e^-$ collisions, which are commonly used as a proxy for the hadronization environment of DM annihilation into quarks.

The tuning strategy adopted here focuses on reproducing antiproton observables over the widest possible energy range. In principle, a complete validation would also require antineutron data. However, (anti)neutrons are notoriously difficult to measure in collider environments and no dedicated \textsf{LHC} measurements exist. A possible hint of an isospin asymmetry, with $\bar n/\bar p \simeq 1.2$--$1.3$ in fixed-target $pp$ at $\sqrt{s}\simeq 17$~GeV, was reported by \textsf{NA49} \cite{Fischer:2002qp}, but it has not been firmly established by subsequent publications. At high energies, inclusive multiplicities of $p$ and $\bar p$ are comparable, as are those of charged and neutral pions \cite{ALICE:2013yba}, suggesting approximate isospin symmetry in soft production. For these reasons, and consistently with the default \texttt{PYTHIA} treatment (i.e.\ the \texttt{Monash} tune~\cite{Skands:2014pea}), we assume equal $\bar p$ and $\bar n$ yields in $pp$ collisions in the remainder of this work.

\begin{figure}
    \centering
    \includegraphics[width=0.99\linewidth]{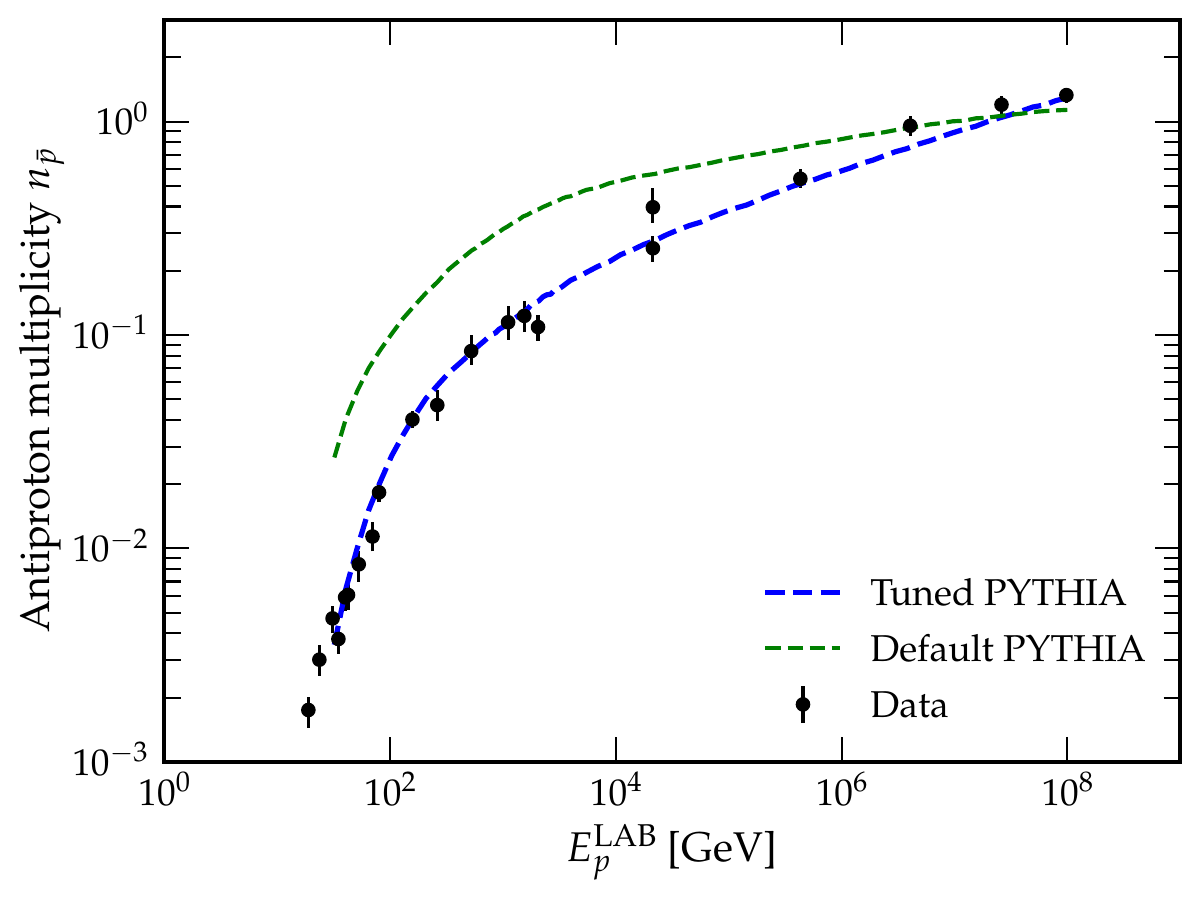}
    \caption{Antiproton multiplicity $n_{\bar p}=N_{\bar p}/N_{\rm ev}$. Antiproton multiplicity in $pp$ collisions as a function of the (equivalent fixed-target) proton beam energy $E_p^{\rm LAB}$, compiled from Refs.~\cite{Antinucci:1972ib,NA61SHINE:2017fne,NA49:2009brx,PHENIX:2011rvu,ALICE:2011gmo,ALICE:2015ial}. Black dots show the experimental measurements. The blue dashed curve is the prediction obtained with our energy-dependent \texttt{PYTHIA}~8.315 tune (see text), while the green dashed curve corresponds to the default \texttt{PYTHIA}~8.315 configuration.}
    \label{fig:antipmult}
\end{figure}

The default \texttt{PYTHIA} configuration does not reproduce satisfactorily the low-energy $\bar p$ production measured in fixed-target experiments (e.g.~\textsf{NA61} and \textsf{NA49}), as illustrated by the green curve in Fig.~\ref{fig:antipmult}. In particular, at $E_p^{\rm LAB}\sim (30$--$200)~\mathrm{GeV}$ the default setup overpredicts $n_{\bar p}$ by up to about an order of magnitude, which would directly bias the predicted antinuclei yields in the region of low kinetic energy per nucleon $K$ most relevant for DM searches. This motivates an energy-dependent tune that remains compatible with low-energy fixed-target data and high-energy collider measurements simultaneously.

We build our setup starting from the \texttt{Tune 2M} baseline \cite{Corke:2010yf}, together with the prescriptions denoted as \texttt{Mode~2} in Ref.~\cite{Christiansen:2015yqa}. On top of that, we introduce an explicit energy dependence for a small set of fragmentation parameters that are known to affect baryon production and the transverse-momentum spectra:

\begin{itemize} 

\item \texttt{StringPT:sigma}. This parameter controls the width of the Gaussian transverse kicks assigned at each string breaking. \texttt{PYTHIA} draws independent Gaussians in $p_x$ and $p_y$ with variance $\sigma^2$, such that $\langle p_T^2\rangle = 2\sigma^2$. We use \begin{eqnarray} && \texttt{StringPT:sigma} = \\ && \begin{cases} 0.40, & E_p^{\rm LAB} \leq 1~{\rm TeV},\\ 0.40+0.035\,\log_{10}\!\left(\dfrac{E_p^{\rm LAB}}{1~{\rm TeV}}\right), & E_p^{\rm LAB} > 1~{\rm TeV}. \nonumber \end{cases} \end{eqnarray} 

\item \texttt{StringFlav:probQQtoQ}. This parameter sets the suppression of diquark production relative to quark production, and thus regulates baryon versus meson production. We adopt \begin{eqnarray} &&\texttt{StringFlav:probQQtoQ} = \\ && \begin{cases} 0.055-0.017\,\log_{10}^2\!\left(\dfrac{E_p^{\rm LAB}}{1~{\rm TeV}}\right), & E_p^{\rm LAB} \leq 1~{\rm TeV},\\[3pt] 0.030+0.002\,\log_{10}\!\left(\dfrac{E_p^{\rm LAB}}{1~{\rm TeV}}\right), & E_p^{\rm LAB} > 1~{\rm TeV}. \nonumber \end{cases} \end{eqnarray} 

\item \texttt{StringFlav:probStoUD}. This parameter controls the suppression of strange-quark production relative to the up and down ones. We set it to $0.217$ for $E_p^{\rm LAB}<1$~TeV and to $0.200$ at higher energies. 

\end{itemize}

Although the tune is energy dependent, we verified that the resulting hadron multiplicities (including pions and baryons) evolve smoothly with energy and do not show visible discontinuities across the transition at $E_p^{\rm LAB}=1$~TeV.

We validate the tune using two complementary classes of measurements:
(i) the \emph{multiplicity} $n_{\bar p} = N_{\bar p}/N_{\rm ev}$ (average number of antiprotons per inelastic $pp$ collision, integrated over transverse momentum and rapidity), which robustly constrains the overall normalization and its energy evolution; and
(ii) differential multiplicity distributions (hereafter referred to as \emph{spectra}), which probe the kinematic shapes relevant for coalescence (in particular as a function of transverse momentum $p_T$, rapidity $y$, or $x_F$).

Fig.~\ref{fig:antipmult} shows $n_{\bar p}$ as a function of the incoming proton energy for both the default \texttt{PYTHIA} configuration and our tune. For collider measurements quoted at a center-of-mass energy $\sqrt{s}$, we report the equivalent fixed-target beam energy
\begin{equation}
E_p^{\rm LAB} = \frac{s-2m_p^2}{2m_p},
\end{equation}
so that fixed-target and collider results can be displayed on the same axis. We consider data from \textsf{NA61}, \textsf{NA49}, \textsf{PHENIX}, \textsf{ALICE} and older measurements \cite{Antinucci:1972ib,NA61SHINE:2017fne,NA49:2009brx,PHENIX:2011rvu,ALICE:2011gmo,ALICE:2015ial}, spanning many orders of magnitude in $E_p^{\rm LAB}$.%
\footnote{\textsf{ALICE} measures $dn_{\bar p}/dy$ at midrapidity, $|y|<0.5$. To compare with full-phase-space multiplicities, we rescale the \textsf{ALICE} results using the rapidity distribution predicted by our tuned \texttt{PYTHIA} configuration.}
The agreement between the data and our energy-dependent tune indicates that the global $\bar p$ yield is under control across the full energy range of interest.

\begin{figure}
    \centering
    \includegraphics[width=0.99\linewidth]{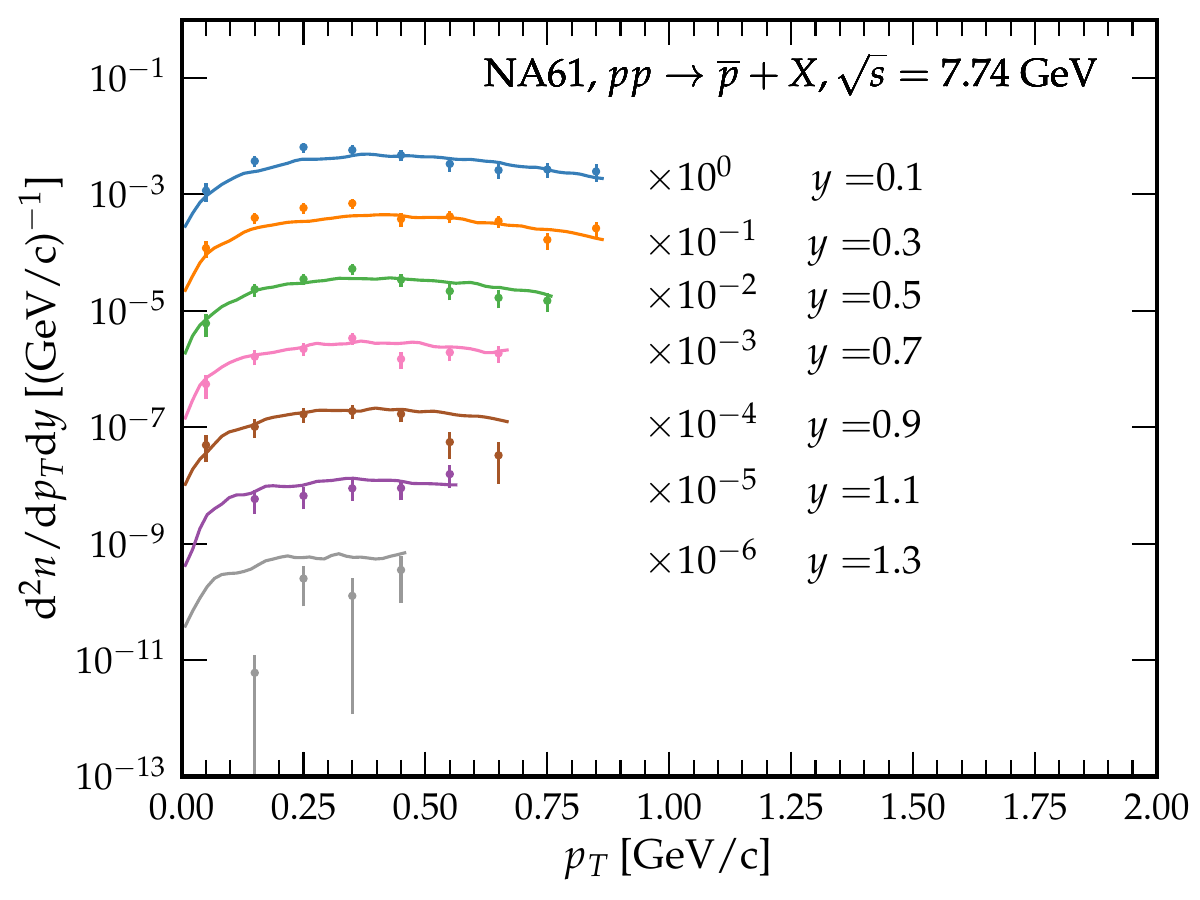}
    \includegraphics[width=0.99\linewidth]{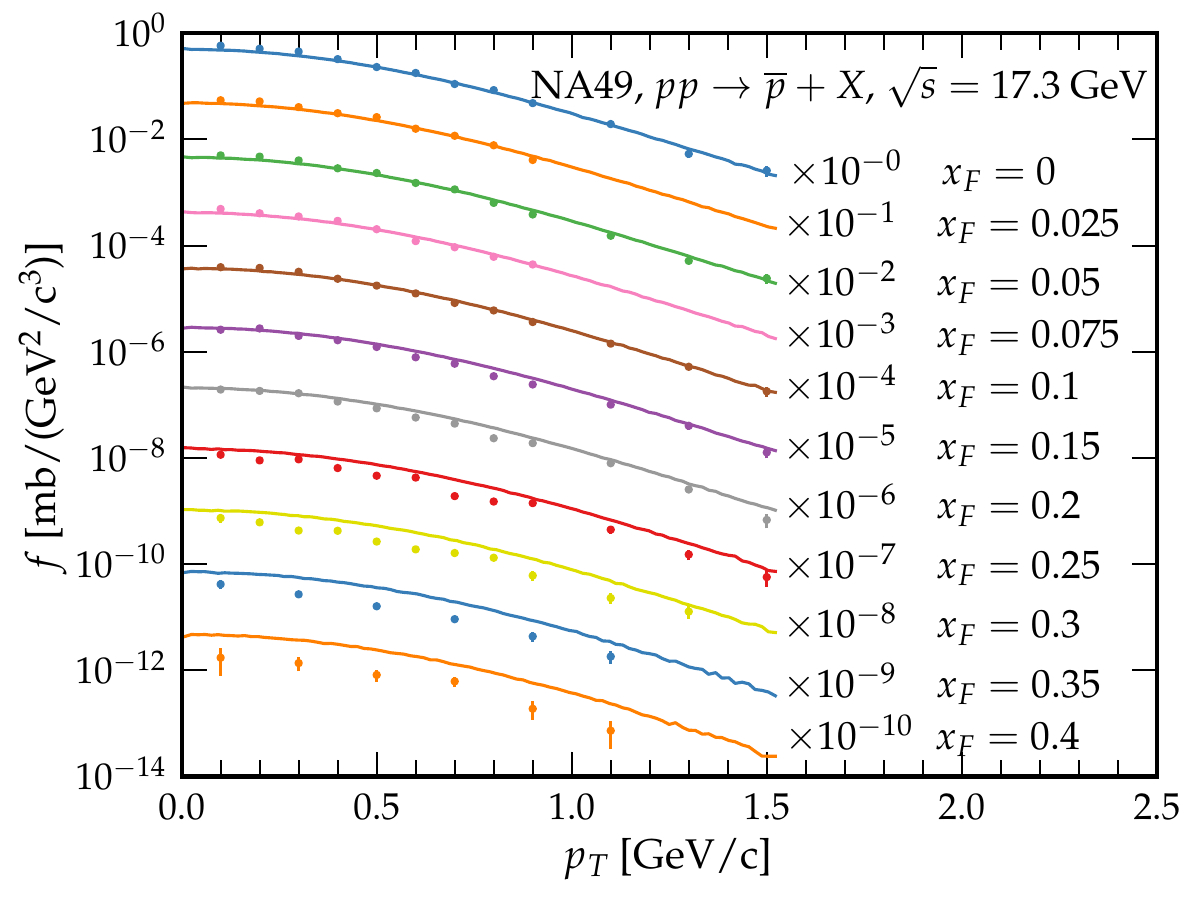}
    \includegraphics[width=0.99\linewidth]{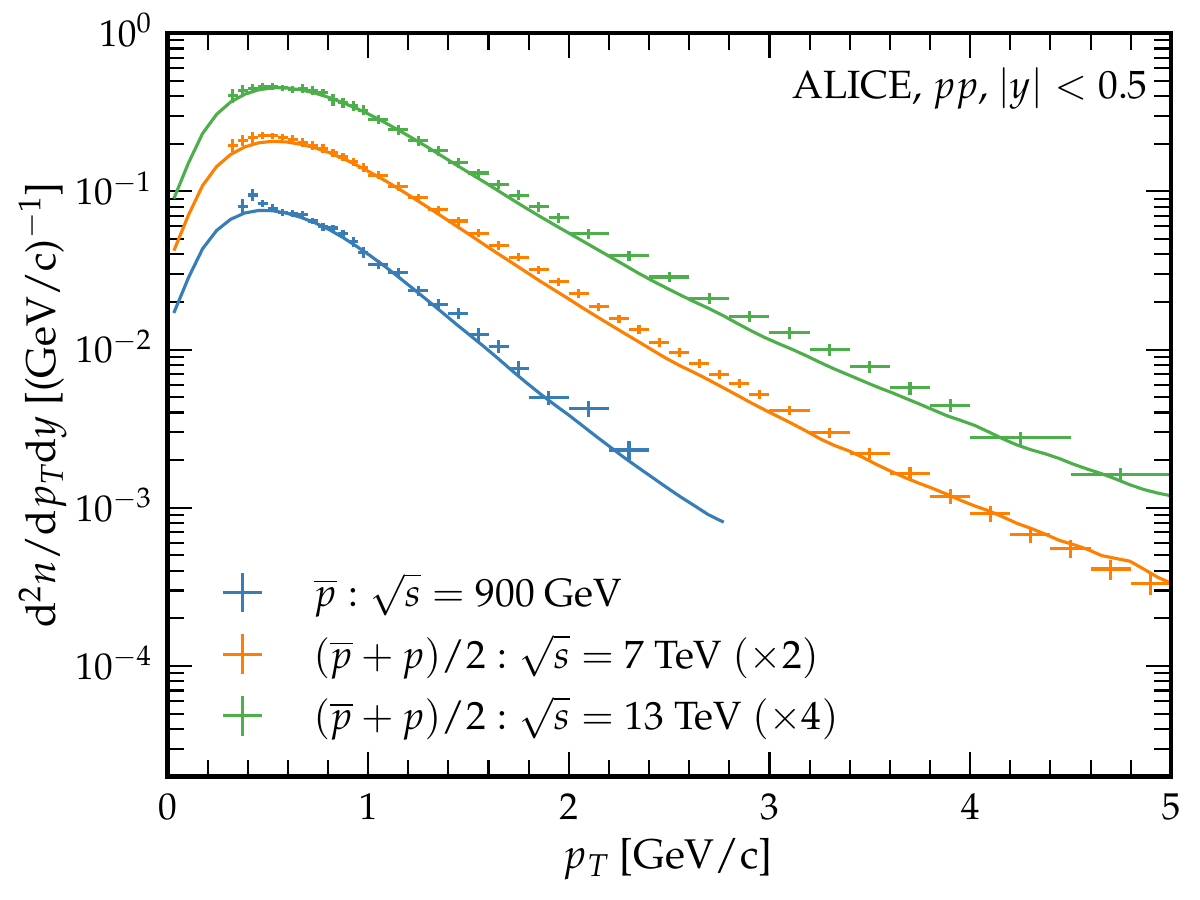}
    \caption{Validation of our \texttt{PYTHIA} tune with antiproton spectra. \emph{Top:} $d^2n/(dp_T\,dy)$ in fixed-target $pp$ collisions at $\sqrt{s}=7.74$~GeV (\textsf{NA61}) for several rapidity bins $y$; spectra are rescaled by the factors indicated in the panel for readability \cite{NA61SHINE:2017fne}. \emph{Middle:} invariant cross section $f=(1/\pi)\,d^3\sigma/(dp_T^2\,dx_F)$ in fixed-target $pp$ at $\sqrt{s}=17.3$~GeV (\textsf{NA49}) for several $x_F$ bins, again rescaled by the indicated powers of ten \cite{NA49:2009brx}. \emph{Bottom:} midrapidity ($|y|<0.5$) $p_T$ spectra in $pp$ collisions measured by \textsf{ALICE} at $\sqrt{s}=0.9$, $7$, and $13$~TeV; the $7$ and $13$~TeV datasets are multiplied by factors $2$ and $4$, respectively, as indicated in the legend \cite{ALICE:2011gmo,ALICE:2015ial,ALICE:2020jsh}. In all panels, the points denote data and solid curves the predictions from our tuned \texttt{PYTHIA} configuration.}
    \label{fig:tuning}
\end{figure}

A  validation on a differential quantity is presented in Fig.~\ref{fig:tuning}. The top panel compares our prediction for $d^2n/(dp_T\,dy)$ to \textsf{NA61} data at $\sqrt{s}=7.74$~GeV \cite{NA61SHINE:2017fne} in several rapidity slices. The middle panel compares the Lorentz-invariant cross section $f=(1/\pi)\,d^3\sigma/(dp_T^2\,dx_F)$ to \textsf{NA49} data at $\sqrt{s}=17.3$~GeV \cite{NA49:2009brx} for a range of $x_F$ values. Here $y$ and $x_F$ trace the longitudinal momentum $p_L$ in the center-of-mass frame,
\begin{equation}
y \equiv \frac12\ln\!\left(\frac{E+p_L}{E-p_L}\right),\qquad
x_F \equiv \frac{2p_L}{\sqrt{s}}.
\end{equation}
We observe that the tune reproduces the overall shapes and normalizations well; residual discrepancies become more visible in the most forward bins (large $y$ or $x_F$), where the experimental uncertainties also increase. Finally, the bottom panel compares our prediction to \textsf{ALICE} midrapidity spectra at $\sqrt{s}=0.9$, $7$, and $13$~TeV \cite{ALICE:2011gmo,ALICE:2015ial,ALICE:2020jsh}, showing good agreement also in the high-energy regime.

In Appendix~\ref{appx:tuning}, we provide additional comparisons for other kinematic projections. Overall, our tune yields a good description of the differential $\bar{p}$ spectra, especially at low $\sqrt{s}$ and low rapidity $y$. We also include comparisons of the $(p+\bar{p})/(\pi^+ + \pi^-)$ ratio, which show some tension at low $p_T$. This discrepancy suggests that the tune does not fully capture the \textsf{NA61} measurements of the charged pion spectra. Even so, our tuning strategy is chiefly intended to model $\bar{p}$ production accurately, so as to provide robust predictions for $\overline{\mathrm{D}}$ production in $pp$ collisions.

\section{PYTHIA setup for antinuclei production}
\label{sec:pythiaDbar}

\subsubsection{Antinuclei spatial and momentum distributions}

In \texttt{PYTHIA}, the transition from partons to hadrons is described by the Lund string model: color confinement is modeled as a relativistic flux tube (a ``string'') stretched between color-connected partons, with an (approximately) constant tension $\kappa \sim \mathcal{O}(1)\,\mathrm{GeV/fm}$. As the partons separate, potential energy builds up in the string until it becomes energetically favorable to break via non-perturbative $q\bar q$ pair creation (often interpreted as tunneling in the string field), producing a sequence of string breakups that are iterated from the endpoints towards the interior. Hadrons are formed by combining adjacent string pieces (and, in baryon production, diquark mechanisms), with longitudinal momentum sharing governed by the Lund fragmentation function and transverse momenta generated from a (roughly) Gaussian kick at each breakup \cite{Andersson:1983ia,Sjostrand:2014zea}. Historically, this framework has been implemented primarily as a momentum-space model, while the space--time structure of hadron formation is necessarily more model dependent; in particular, production ``vertices'' should be understood as effective points inferred from the string worldsheet rather than directly observable formation points of extended bound states \cite{FerreresSole:2018vla}. A concrete space--time prescription for multiparton strings, including practical vertex definitions for hadrons formed between two adjacent string breaks, is provided in \texttt{PYTHIA} following Ref.~\cite{FerreresSole:2018vla}.

The event record can be augmented with space--time information through two conceptually distinct switches:
\begin{itemize}
\item \texttt{PartonVertex:setVertex = on} assigns production vertices to partons from the hard process and the parton shower, using formation-time/virtuality arguments such that these vertices typically lie on femtometer scales. If this option is enabled \emph{without} enabling hadron-vertex generation, the produced hadrons inherit the space--time positions of their parent partons or string endpoints, and thus remain at $\mathcal{O}(1\,\mathrm{fm})$ scales.

\item \texttt{Fragmentation:setVertices = on} activates the Lund-model space--time construction of hadron production points from the string fragmentation process itself, i.e.\ hadron vertices are computed from the space--time locations of the relevant adjacent string breaks as in Ref.~\cite{FerreresSole:2018vla}.
\end{itemize}

Because long strings in boosted topologies can extend significantly along the beam direction, hadron production vertices can reach much larger separations in the pair reference frame (especially in $z$), as illustrated, e.g., by the distributions shown in Fig.~4 of Ref.~\cite{Bierlich:2021whr}. Since our goal is to use \emph{hadron} production vertices (rather than inherited parent positions), we enable both switches.

\begin{figure}
    \centering
    \includegraphics[width=1\linewidth]{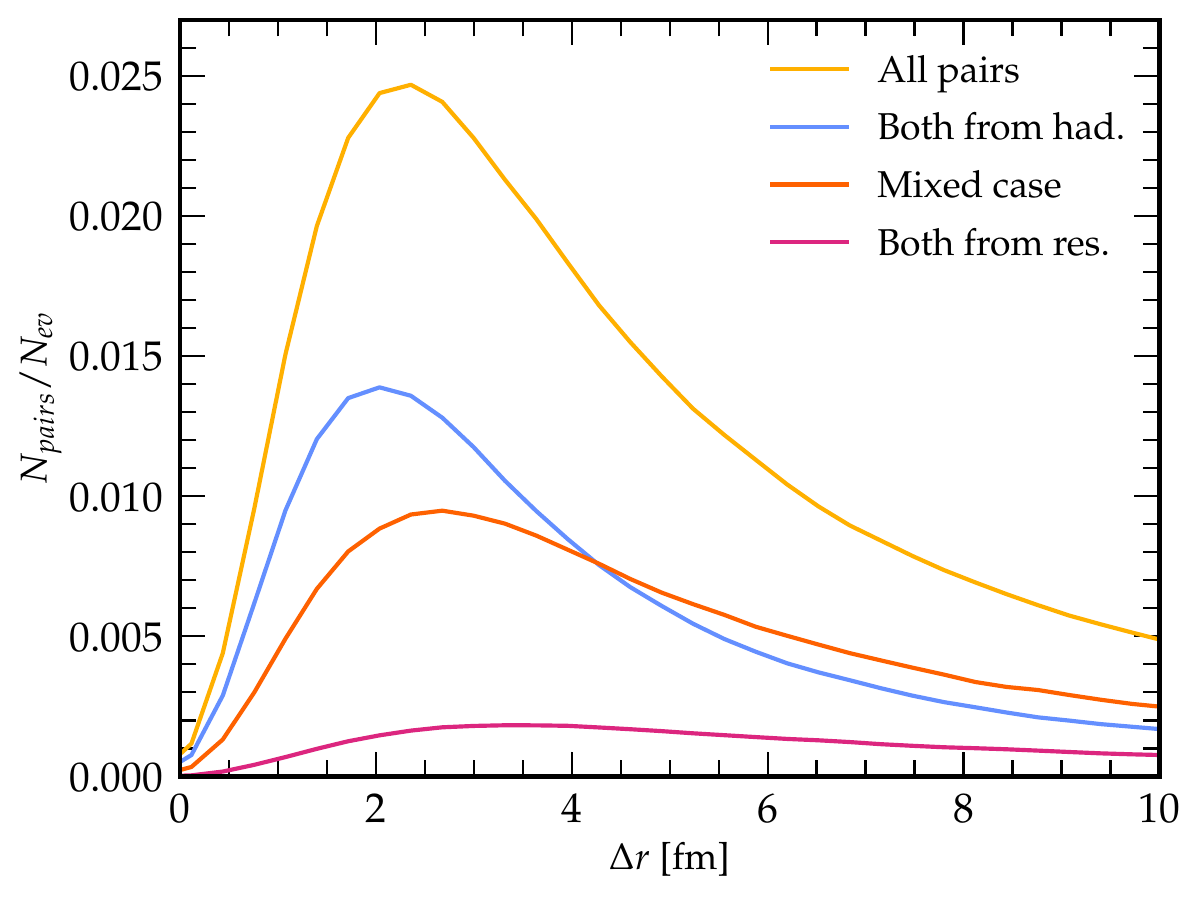}
    \caption{Normalized distribution of the $\bar{p}$--$\bar{n}$ pair separation $\Delta r$, in the pair reference frame, for antinucleons generated with our tuned \texttt{PYTHIA} setup in $pp$ collisions at $\sqrt{s}=7$~TeV. Contributions are shown separately for pairs in which both antinucleons originate from resonance decays (magenta), both originate from hadronization (blue), one originates from hadronization and the other from a resonance decay (orange), and for the inclusive sample of all $\bar p$--$\bar n$ pairs (yellow).
    }
    \label{fig:sourceDbar}
\end{figure}

Figure~\ref{fig:sourceDbar} shows the distribution of spatial separations $|\Delta \vec r|$ between $\bar p$--$\bar n$ pairs, evaluated in the pair rest frame.\footnote{Here $\Delta\vec r\equiv \vec r_{\bar p}-\vec r_{\bar n}$ is constructed from the production vertices stored in the \texttt{PYTHIA} event record.} Pairs produced directly at hadronization peak at $\Delta r\simeq 2~\mathrm{fm}$, whereas pairs in which both antinucleons originate from resonance decays peak at larger separations, $\Delta r\simeq 3.5~\mathrm{fm}$. The hadronization component is also noticeably narrower than the resonance-decay component. This implies that coalescence is generically less efficient for antinucleons produced via resonance decays, since they are typically emitted further apart. The inclusive distribution peaks at $\Delta r\simeq 2.5~\mathrm{fm}$. We also note that the resonance-decay contribution is subdominant in \texttt{PYTHIA} compared to direct hadronization (see Sec.~\ref{sec:pythiaDbar} below).

\begin{figure}
    \centering
    \includegraphics[width=1\linewidth]{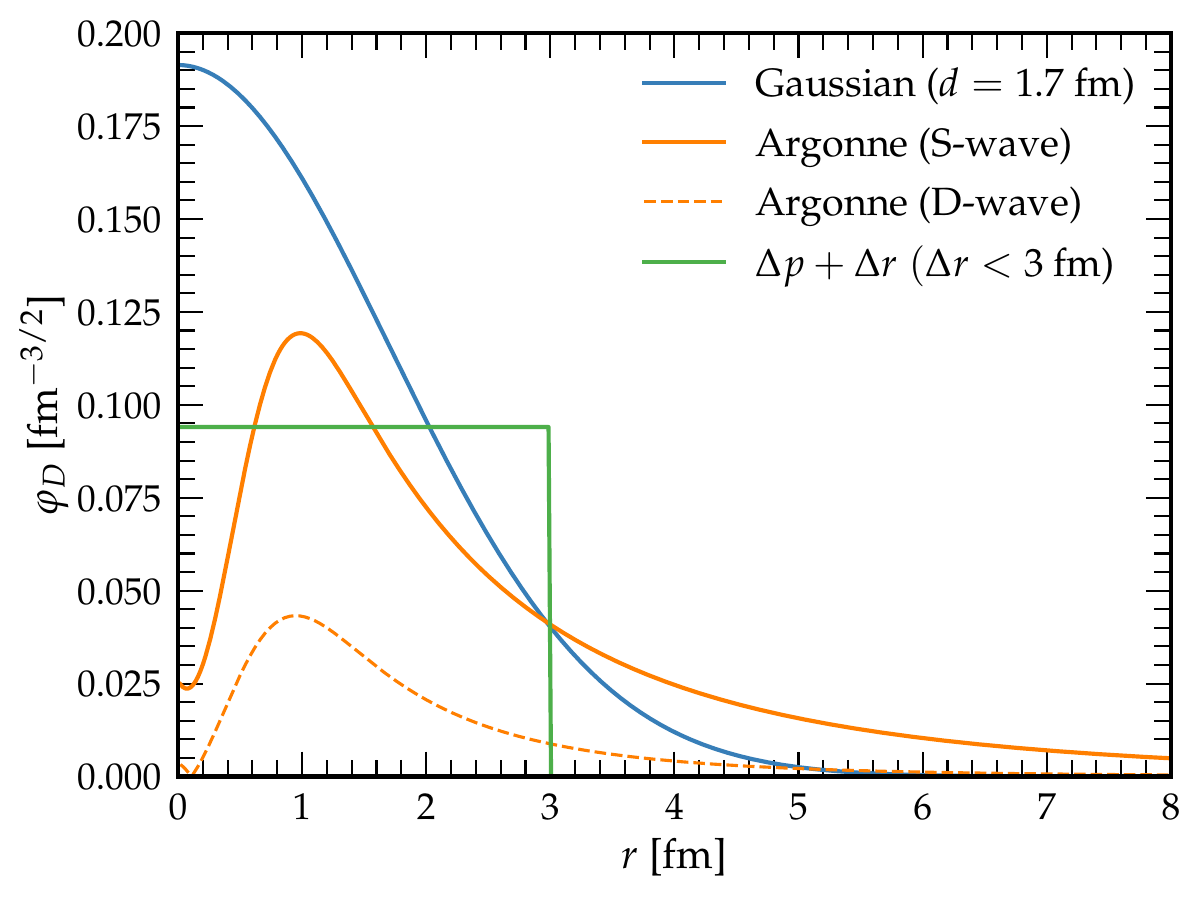}
    \caption{Deuteron wavefunctions $\varphi_{\rm D}(r)$ used in the Wigner-based coalescence models: Gaussian (blue) and Argonne $v_{18}$~\cite{Wiringa:1994wb} S-wave and D-wave components (solid and dashed orange, respectively). The step-like profile corresponding to the $\Delta p+\Delta r$ model is also shown (green) for illustrative purposes, although it is not a Wigner-based prescription.}
    \label{fig:DeuteronWF}
\end{figure}

The antinucleon separation distribution can be compared to the shapes of the deuteron wavefunctions entering the Wigner kernels. The Gaussian and Argonne wavefunctions, shown in Fig.~\ref{fig:DeuteronWF}, are concentrated within the inner few femtometers; the Gaussian is broader than the Argonne one. In particular, about 60\% of the Argonne probability density is contained within $r<3~\mathrm{fm}$. As a result, the Argonne kernel preferentially selects pairs produced at small separations, which in \texttt{PYTHIA} are more often associated with prompt hadronization, whereas the broader Gaussian kernel has a comparatively larger overlap with the resonance-decay tail.

\subsubsection{Resonance and hadronic production of antinuclei}

Recently, the \textsf{ALICE} Collaboration reported that approximately 80\% of (anti-)deuterons are formed through the coalescence of at least one (anti-)nucleon originating from the decay of short-lived resonances~\cite{ALICE:2025byl}. In contrast, \texttt{PYTHIA} produces a significantly smaller fraction of (anti)nucleons from resonance decays. Using both the default \texttt{PYTHIA} settings and our $\bar p$-tuned configuration (which does not modify the resonance production mechanism), we find that about 34\% (22\%) of antiprotons originate from resonance decays in $pp$ collisions at $\sqrt{s}=7$~TeV ($17.3$~GeV).

This discrepancy can be partially attributed to the fact that \texttt{PYTHIA} underestimates the multiplicities of strongly decaying resonances and does not include the full set of resonances listed by the PDG. This effect can be exacerbated when multi-parton interactions and color reconnection are enabled, as in our setup \cite{Acconcia:2017bjv}. In Table~\ref{tab:resonancemult} we compare the predicted multiplicities of several resonances---$\Delta(1232)^{++}$, $\Xi(1530)^0$, and $\Sigma(1385)^+$---with measurements performed by the \textsf{DELPHI}~\cite{DELPHI:1995ysj,DELPHI:2005zmk} and \textsf{OPAL}~\cite{OPAL:1995otk,OPAL:1996gsw} Collaborations in $e^+e^-$ collisions at the $Z$ pole, as reported by the \textsf{PDG}~\cite{ParticleDataGroup:2024cfk}. Similarly, in Table~\ref{tab:resonancemultALICE} we compare the predicted multiplicities of $\Xi(1530)^0$, $\Sigma(1385)^\pm$, and their antiparticles with midrapidity measurements performed by \textsf{ALICE} in $pp$ collisions at $\sqrt{s}=7$~TeV~\cite{ALICE:2014zxz}. In both cases, the \texttt{PYTHIA} predictions are systematically lower than the data, which at least partially explains why \texttt{PYTHIA} underestimates the resonance-decay contribution to (anti)nucleon production.

\begin{table}[t!]
    \centering
    \begin{tabular}{|c|c|c|}
         \hline
         Resonance & \textsf{DELPHI} \& \textsf{OPAL} & \texttt{PYTHIA} \\
         \hline
         $\Delta(1232)^{++}$ & $0.087\pm0.033$ & $0.059$ \\
         \hline
         $\Xi(1530)^0$ & $0.0059\pm0.0011$ & $0.002$ \\
         \hline
         $\Sigma(1385)^+$ & $0.0239\pm0.0014$ & $0.012$ \\
         \hline
    \end{tabular}
    \caption{Comparison between \textsf{DELPHI} and \textsf{OPAL} measured \cite{ParticleDataGroup:2024cfk} and predicted multiplicities of selected resonances in $e^+e^-$ collisions at the $Z$ pole.}
    \label{tab:resonancemult}
\end{table}

\begin{table}[t!]
    \centering
    \begin{tabular}{|c|c|c|}
         \hline
         Resonance & \textsf{ALICE} & \texttt{PYTHIA} \\
         \hline
         $\Sigma(1385)^+$ & $\left(1.00^{+0.15}_{-0.14}\right)\times 10^{-2}$ & $\left(3.30\pm0.06\right)\times 10^{-3}$ \\
         \hline
         $\overline{\Sigma}(1385)^+$ & $\left(1.03^{+0.17}_{-0.15}\right)\times 10^{-2}$ & $\left(3.16\pm0.06\right)\times 10^{-3}$ \\
         \hline
         $\Sigma(1385)^-$ & $\left(1.08^{+0.17}_{-0.16}\right)\times 10^{-2}$ & $\left(3.04\pm0.06\right)\times 10^{-3}$ \\
         \hline
         $\overline{\Sigma}(1385)^-$ & $\left(0.91^{+0.15}_{-0.14}\right)\times 10^{-2}$ & $\left(2.91\pm0.06\right)\times 10^{-3}$ \\
         \hline
         $\Xi(1530)^0$ & $\left(2.56^{+0.41}_{-0.38}\right)\times 10^{-3}$ & $\left(6.7\pm0.3\right)\times 10^{-4}$ \\
         \hline
    \end{tabular}
    \caption{Comparison between \textsf{ALICE} measured \cite{ALICE:2025byl} and predicted multiplicities of selected resonances in $pp$ collisions at midrapidity ($|y|<0.5$).}
    \label{tab:resonancemultALICE}
\end{table}

Several models of hadron production in high-energy collisions, in particular hadron resonance gas (HRG) models that treat hadrons as an ideal gas, predict a substantially larger fraction of nucleons originating from resonance decays, in closer agreement with the \textsf{ALICE} results than \texttt{PYTHIA}. Using the \texttt{Thermal-FIST} package~\cite{Vovchenko:2019pjl}, we find that in $pp$ collisions at $\sqrt{s}=13$~TeV, approximately 73\% of nucleons are produced via resonance decays.

Since this result is directly constrained by experimental data only at \textsf{LHC} energies, its extension to lower collision energies remains speculative in the absence of additional measurements. Nevertheless, if HRG models provide an accurate description of hadron production in high-energy collisions, one would expect the fraction of nucleons originating from resonance decays to be only weakly dependent on the collision energy. Consistent with this expectation, \texttt{Thermal-FIST} yields a similar ratio for $pp$ collisions at $\sqrt{s}=900$~GeV.

In an attempt to reproduce a comparable fraction of nucleons from resonance decays within \texttt{PYTHIA}, we explored discarding a large fraction of $pp$ collision events in which antiprotons are produced directly during hadronization. This procedure naturally reduces the total $\bar p$ multiplicity, since \texttt{PYTHIA} predominantly produces nucleons through hadronization rather than resonance decays. To compensate for this effect, we increased the overall baryon production by tuning the \texttt{StringFlav:probQQtoQ} parameter. However, reproducing the experimentally measured $\bar p$ multiplicity would require setting this parameter beyond its maximum value allowed in the code. Consequently, within the current version of \texttt{PYTHIA}, it is not possible to reproduce the resonance-decay fraction reported by \textsf{ALICE} without introducing tensions with measurements of the total $\bar p$ multiplicity. For this reason, we retain our $\bar p$-tuned \texttt{PYTHIA} configuration as the benchmark setup for the remainder of this paper.

\section{Determining the coalescence parameters}
\label{sec:results}

\subsection{Fitting \textsf{ALICE} (anti)deuteron spectra}
\label{subsec:fit_alice}

To calibrate the coalescence-model parameters we perform a simultaneous fit to the \textsf{ALICE} $p_T$-differential spectra of deuterons and antideuterons measured at
$\sqrt{s}=0.9,\ 2.76,\ 7$~TeV \cite{ALICE:2017xrp} and $\sqrt{s}=13$~TeV \cite{ALICE:2020foi}.
All spectra are provided at midrapidity, $|y|<0.5$.
We then compare the resulting coalescence parameters with those extracted from the \textsf{ALEPH} $\overline{\mathrm D}$ multiplicity in hadronic $Z$ decays, as reported in Ref.~\cite{DiMauro:2024kml}, to test the universality of the coalescence description.
We apply the analysis to the $\Delta p$, $\Delta p+\Delta r$ and {\tt Gaussian} implementations. Since the {\tt Argonne} setup does not contain any free parameter, as its wavefunction is fully fixed by nucleon–nucleon scattering data
and the deuteron binding energy, we therefore only show its predictions.

As discussed in Ref.~\cite{DiMauro:2024kml}, coalescence prescriptions that include one parameter controlling the spatial distribution and one controlling the momentum dependence can exhibit a strong degeneracy. In the {\tt Gaussian} setup, for instance, changes in the effective source size $\sigma$ can be partially compensated by changes in the bound-state scale (here denoted by $\delta$), leading to families of parameter pairs with similar goodness of fit.
To reduce this degeneracy and keep the comparison across datasets and collision systems transparent, we fix the spatial scale to a benchmark value of $\sigma = 3~\mathrm{fm}$, comparable to the deuteron size inferred from its RMS charge radius \cite{CREMA:2016idx}.
Concretely, we adopt:
(i) a single free parameter $p_{\rm coal}$ for the $\Delta p$ model;
(ii) a single free parameter $p_{\rm coal}$ for the $\Delta p+\Delta r$ model, fixing $r_{\rm coal}\equiv \Delta r = 3~\mathrm{fm}$;
(iii) a single free parameter $\delta$ for the {\tt Gaussian} model, fixing the source size $\sigma=3~\mathrm{fm}$.
The {\tt Argonne} implementation is treated as parameter free and is therefore not tuned to the data.

The baryon-emission source radius inferred from \textsf{ALICE} femtoscopy measurements in high-multiplicity $pp$ collisions provides an important qualitative reference for the space-time structure relevant to antinuclei formation; however, since its extraction depends on the assumed source function, the treatment of final-state interactions, and corrections for resonance decays, it cannot be directly interpreted as a universal coalescence radius applicable to all event-by-event coalescence models~\cite{Acharya:2020source}.

We generate \texttt{PYTHIA} simulations of $pp$ collisions to obtain a pool of antinucleons. For each coalescence model, we then scan a discrete grid of the relevant free parameter, applying the coalescence criteria to the same set of events. This procedure minimizes relative statistical fluctuations and isolates the dependence on the coalescence parameter. We compute the predicted spectra in the same kinematic bins as \textsf{ALICE} and evaluate a global $\chi^2$ over all energies and over both D and $\dbar$ datasets. We then interpolate $\chi^2$ as a function of the scanned parameter to extract the best-fit value and its $1\sigma$ uncertainty.
We simulated 100, 150, 450, and 450  million events for $\sqrt{s} = 0.9$, 2.76, 7, and 13~TeV, respectively. These sample sizes ensure that the Monte-Carlo relative statistical uncertainty remains below $5\%$ for each experimental data point in $p_T$, for the parameter point closest to the best fit. For example, for the 7 and 13~TeV cases, the relative statistical uncertainty is $\approx 1\%$ in the lowest $p_T$ bins and stays below $5\%$ even in the highest $p_T$ bins included in our analysis.

\begin{figure}
    \centering
    \includegraphics[width=1\linewidth]{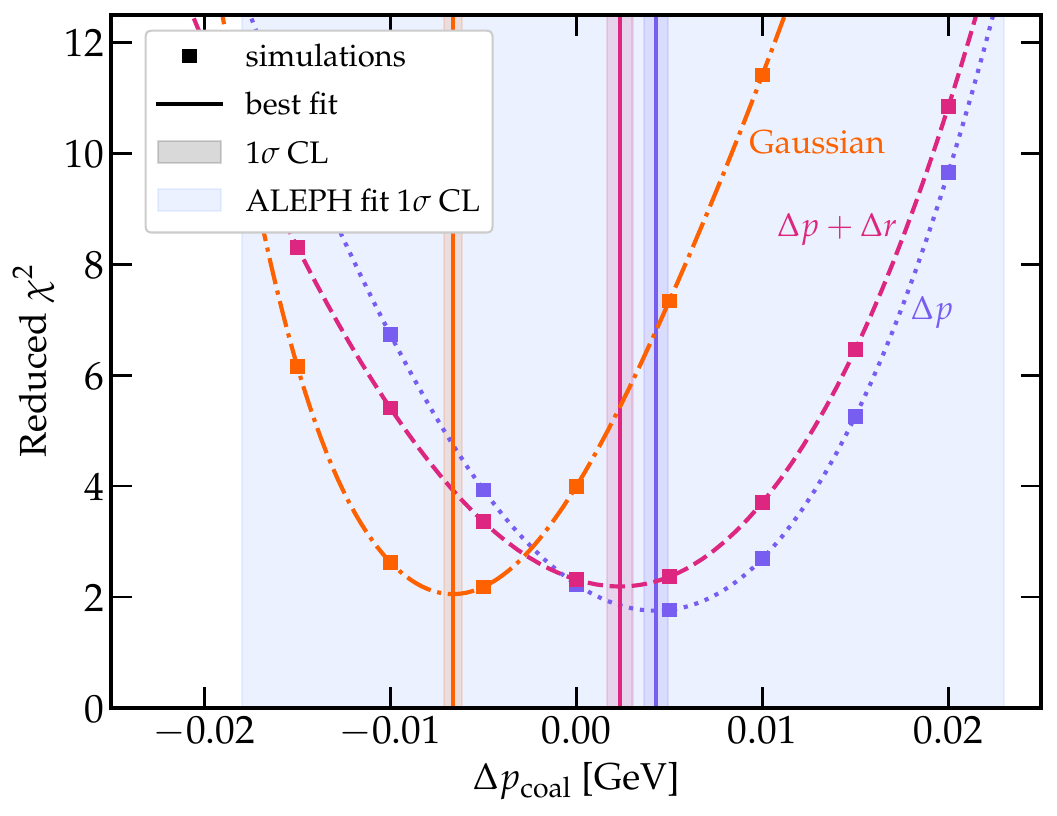}
    \caption{Reduced $\chi^{2}$ profiles profiles as a function of $\Delta p_{\rm coal}\equiv p_{\rm coal}-p_{\rm coal}^{\rm ALEPH}$, where $p_{\rm coal}^{\rm ALEPH}$ is the best-fit for the coalescence parameter found by fitting ALEPH data as reported in Ref.~\cite{DiMauro:2024kml}, while $p_{\rm coal}$ is the best-fit obtained by fitting \textsf{ALICE} deuteron and antideuteron spectra measured at $\sqrt{s} = 0.9, 2.76, 7$, and $13$~TeV. The squared markers indicate the $\chi^2$ obtained when simulating the $D$ and $\overline{D}$ data at specific values of $p_{\rm{coal}}$.
    The curves represent the interpolations of the single $\chi^2$ points.
    We show the results for the following models: $\Delta p$ (dotted purple), $\Delta p+\Delta r$ (dashed magenta), and {\tt Gaussian} (dash-dotted orange). In the {\tt Gaussian} model we rescaled the $\delta$ parameter by a factor $0.11\,\mathrm{GeV}/1.80\,\mathrm{fm}$. For each model, the vertical solid line and the correspondingly colored shaded band denote the extracted best-fit value and its $1\sigma$ interval, respectively. Additionally, the blue shaded band represents the $1\sigma$ interval obtained from the \textsf{ALEPH} fit for the $\Delta p$ case, plotted for visual comparison. }
    \label{fig:Chi_2_total}
\end{figure}

\begin{table*} 
    \centering
    \renewcommand{\arraystretch}{1.2} 
    \caption{Comparison of coalescence parameters extracted from $pp$ and $e^+e^-$ collisions. The \textsf{ALICE} results correspond to a simultaneous global fit of deuteron and antideuteron spectra across four center-of-mass energies ($\sqrt{s} = 0.9, 2.76, 7, 13$~TeV). The \textsf{ALEPH} results refer to antideuteron production in $e^+e^-$ collisions at the $Z$ resonance ($\sqrt{s} = 91$~GeV) and are taken from Ref.~\cite{DiMauro:2024kml}.}
    \label{tab:fit_comparison_total}
    \begin{tabular*}{\textwidth}{@{\extracolsep{\fill}}l c c c} 
        \toprule
        \textbf{Model} & \textbf{\textsf{ALICE} (Global fit)} & \boldmath{$\chi^2/N_{\text{dof}}$} & \textbf{\textsf{ALEPH} (Reference)} \\
         & \footnotesize{$pp$ @ $0.9$--$13$ TeV} & \footnotesize{(\textsf{ALICE})} & \footnotesize{$e^+e^-$ @ $91$ GeV} \\
        \midrule
        $\Delta p$ & $p_{\text{coal}} = 0.201 \pm 0.001$ GeV & 1.75 & $p_{\text{coal}} = 0.196^{+0.018}_{-0.023}$ GeV \\
        $\Delta p+\Delta r$ & $p_{\text{coal}} = 0.214 \pm 0.001$ GeV & 2.19 & $p_{\text{coal}} = 0.212^{+0.019}_{-0.024}$ GeV \\
        {\tt Gaussian} & $\delta = 1.69 \pm 0.01$ fm & 2.05 & $\delta = 1.80^{+0.2}_{-0.3}$ fm \\
        \bottomrule
    \end{tabular*}
\end{table*}

\begin{table*}
    \centering
    \renewcommand{\arraystretch}{1.2}
    \caption{Fit quality excluding the highest-energy dataset. Same comparison as in Table~\ref{tab:fit_comparison_total}, but performing the \textsf{ALICE} global fit only for $\sqrt{s} = 0.9, 2.76,$ and $7$~TeV (thus excluding 13~TeV).}
    \label{tab:fit_comparison_no_13TeV}
    \begin{tabular*}{\textwidth}{@{\extracolsep{\fill}}l c c c} 
        \toprule
        \textbf{Model} & \textbf{\textsf{ALICE} (Lower $\sqrt{s}$ only)} & \boldmath{$\chi^2/N_{\text{dof}}$} & \textbf{\textsf{ALEPH} (Reference)} \\
         & \footnotesize{$pp$ @ $0.9$--$7$ TeV} & \footnotesize{(\textsf{ALICE})} & \footnotesize{$e^+e^-$ @ $91$ GeV} \\
        \midrule
        $\Delta p$ & $p_{\text{coal}} = 0.198 \pm 0.001$ GeV & 0.73 & $p_{\text{coal}} = 0.196^{+0.018}_{-0.023}$ GeV \\
        $\Delta p+\Delta r$ & $p_{\text{coal}} = 0.211 \pm 0.001$ GeV & 0.91 & $p_{\text{coal}} = 0.212^{+0.019}_{-0.024}$ GeV \\
        {\tt Gaussian} & $\delta = 1.72 \pm 0.01$ fm & 0.75 & $\delta = 1.80^{+0.2}_{-0.3}$ fm \\
        \bottomrule
    \end{tabular*}
\end{table*}

Fig.~\ref{fig:Chi_2_total} shows the $\chi^{2}$ profiles resulting from the \textsf{ALICE} global fit. The derived best-fit values for the free parameter of each model are reported in Table~\ref{tab:fit_comparison_total}, which compares them to the corresponding values extracted from \textsf{ALEPH} $e^+e^-$ data in Ref.~\cite{DiMauro:2024kml}.
Thanks to the small experimental uncertainties of the \textsf{ALICE} spectra and the large Monte-Carlo statistics, the \textsf{ALICE}-derived parameters $p_{\rm coal}$ and $\delta$ are determined at the sub-percent level (relative uncertainties $\sim 0.5\%$).
By contrast, the \textsf{ALEPH}-based calibration in Ref.~\cite{DiMauro:2024kml} yields $\mathcal{O}(10\%)$ relative uncertainties, driven primarily by the larger experimental error on the measured $\overline{\mathrm D}$ multiplicity ($\sim 15$--$20\%$).

We find good overall consistency between the parameters inferred from \textsf{ALICE} $pp$ data and from hadronic $Z$ decays.
While the global-fit $\chi^2/N_{\rm dof}$ values in Table~\ref{tab:fit_comparison_total} are of order $1.7$--$2.2$, Table~\ref{tab:fit_comparison_no_13TeV} shows that excluding the $\sqrt{s}=13$~TeV data significantly improves the fit quality ($\chi^2/N_{\rm dof}\sim 0.73$--$0.91$) while shifting the best-fit parameters only mildly.
This indicates that the extracted coalescence scales are stable, and that the residual tension is mainly associated with the highest-energy dataset (or, equivalently, with the generator-level description of the input (anti)nucleon kinematics at 13~TeV rather than with the coalescence prescription itself).

For the momentum-cut model we obtain $p_{\rm coal}\simeq 0.20$~GeV, fully compatible with the \textsf{ALEPH} determination within uncertainties.
When adding the spatial requirement ($\Delta p+\Delta r$), the best fit shifts to a slightly larger value, $p_{\rm coal}\simeq 0.21$~GeV.
This is expected: imposing $\Delta r<3$~fm removes candidate $\bar p$--$\bar n$ pairs that would otherwise satisfy the momentum criterion only, and a larger $p_{\rm coal}$ compensates this reduction to reproduce the same  $\overline{\rm D}$ yield.

The agreement of the extracted parameters across (i) different production environments (hadronic $pp$ at the \textsf{LHC} versus hadronic fragmentation in $e^+e^-$ at the $Z$ pole) and (ii) widely separated energies ($\sqrt{s}=91$~GeV versus $\sqrt{s}\gtrsim 0.9$~TeV) supports a \emph{universal} coalescence description at the level probed by current collider data. This universality is a key ingredient to reduce the source-model uncertainty in cosmic antinuclei predictions.

\begin{figure}
    \centering
    \includegraphics[width=1\linewidth]{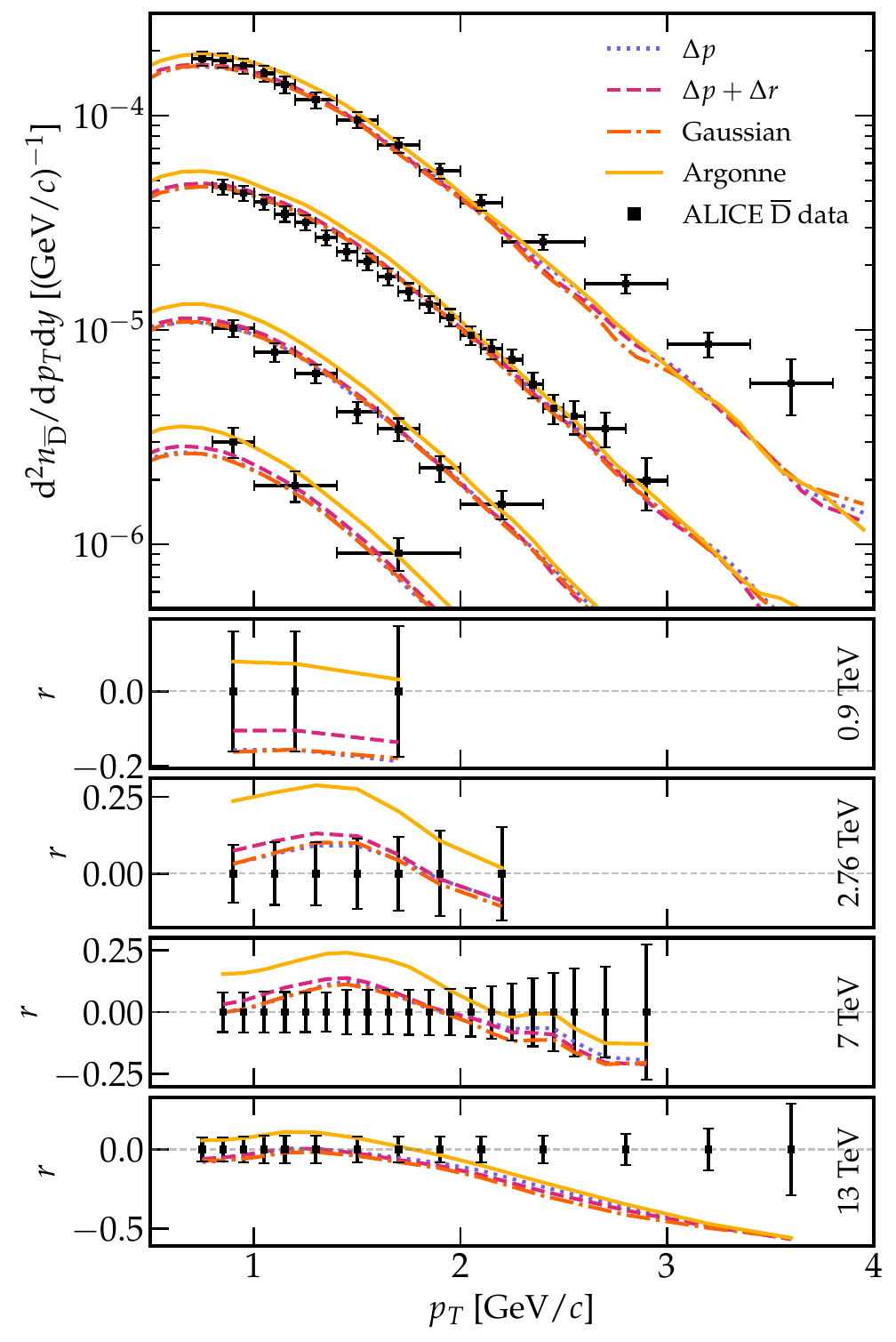}
    \caption{Antideuteron $p_T$-differential spectra. The upper panel  shows the \textsf{ALICE} $\overline{\mathrm D}$ spectra at midrapidity ($|y|<0.5$) for $\sqrt{s}=13,\ 7,\ 2.76,\ 0.9$~TeV (from top to bottom, compared to best-fit predictions of the tuned coalescence models ($\Delta p$, $\Delta p+\Delta r$, {\tt Gaussian}) and to the parameter-free {\tt Argonne} implementation (lines). For readability, the spectra (data and curves) are rescaled by factors $1$, $1/3$, $1/10$, and $1/28$, respectively. The four lower panels show the relative deviation $r=(\mathrm{model}-\mathrm{data})/\mathrm{data}$ for each energy.}
    \label{fig:all_energies_diff_mult_Dbar}
\end{figure}

\begin{figure}
    \centering
    \includegraphics[width=1\linewidth]{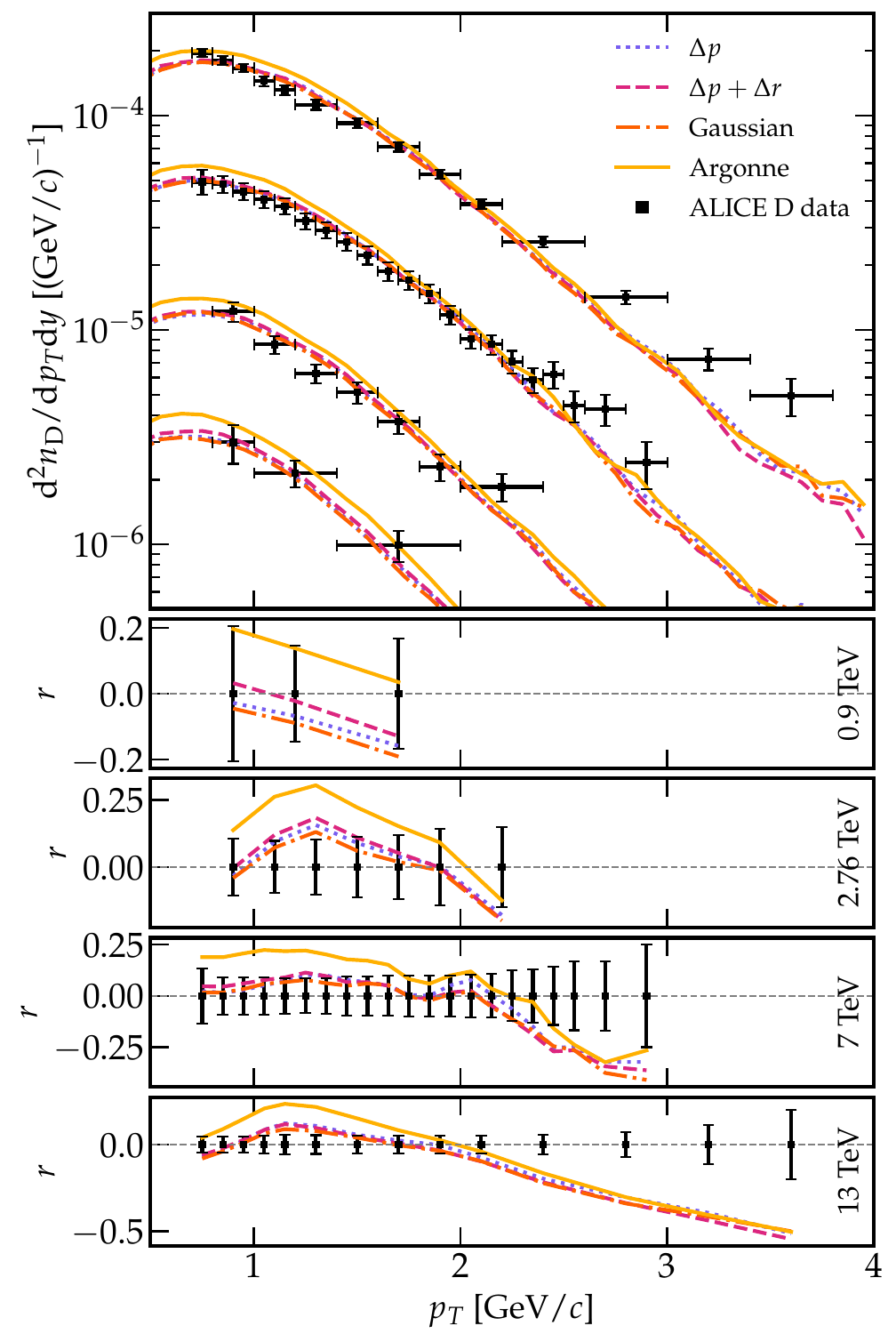}
    \caption{Deuteron $p_T$-differential spectra.
    Same as Fig.~\ref{fig:all_energies_diff_mult_Dbar}, but for the \textsf{ALICE} D spectra at midrapidity.}
    \label{fig:all_energies_diff_mult_D}
\end{figure}

In Figs.~\ref{fig:all_energies_diff_mult_Dbar} and \ref{fig:all_energies_diff_mult_D} we compare the \textsf{ALICE} data with the best-fit predictions for $\overline{\rm D}$ and ${\rm D}$, respectively.
The three tuned setups ($\Delta p$, $\Delta p+\Delta r$, {\tt Gaussian}) yield very similar differential multiplicities, typically differing by $\lesssim 10\%$ over the kinematic range covered by the data. This mirrors what was observed in the \textsf{ALEPH}-based calibration \cite{DiMauro:2024kml}: once tuned to a reference dataset, different reasonable coalescence implementations converge to nearly indistinguishable $\overline{\rm D}$ spectra. For completeness, in Tab.~\ref{tab:d_dbar_integrated_yield_comparison_total} we provide the differential multiplicities $dn/dy$ of (anti-)deuterons at midrapidity obtained in our simulations compared to \textsf{ALICE} data.

\begin{table*}[htbp]
    \centering
    \renewcommand{\arraystretch}{1.2}
    \caption{Midrapidity yields of deuterons and antideuterons.
    Differential multiplicities $dn/dy$ at midrapidity ($|y|<0.5$) extracted from the simulations for the considered coalescence models, compared to \textsf{ALICE} minimum-bias data at $\sqrt{s} = 0.9$, 2.76, 7~\cite{ALICE:2017xrp}, and 13~TeV~\cite{ALICE:2020foi}. Values are scaled by a factor $10^4$ for readability.}
    \label{tab:d_dbar_integrated_yield_comparison_total}
    \begin{tabular*}{\textwidth}{@{\extracolsep{\fill}}l c c c c} 
        \toprule
        \textbf{Data and models} & \textbf{0.9 TeV} & \textbf{2.76 TeV} & \textbf{7 TeV} & \textbf{13 TeV}\\
        \midrule
        \multicolumn{5}{c}{\textbf{Antideuterons ($\overline{\mathrm{D}}$), $dn/dy \times 10^{4}$}} \\
        \midrule
        \textsf{ALICE} & $1.11 \pm 0.10 \pm 0.09$ & $1.37 \pm 0.04 \pm 0.12$ & $1.92 \pm 0.02 \pm 0.15$ & $2.54 \pm 0.03 \pm 0.30$\\
        $\Delta p$ & $0.933 \pm 0.005$ & $1.434 \pm 0.009$ & $1.980 \pm 0.008$ & $2.46 \pm 0.01$\\
        $\Delta p+\Delta r$ & $0.994 \pm 0.005$ & $1.480 \pm 0.009$ & $2.020 \pm 0.008$ & $2.48 \pm 0.01$ \\
        {\tt Gaussian} & $0.928 \pm 0.005$ & $1.437 \pm 0.008$ & $1.959 \pm 0.008$ & $2.42 \pm 0.01$\\
        {\tt Argonne} & $1.208 \pm 0.005$ & $1.713 \pm 0.009$ & $2.252 \pm 0.009$ & $2.74 \pm 0.01$\\
        \midrule
        \multicolumn{5}{c}{\textbf{Deuterons ($\mathrm{D}$), $dn/dy \times 10^{4}$}} \\
        \midrule
        \textsf{ALICE} & $1.12 \pm 0.09 \pm 0.09$ & $1.53 \pm 0.05 \pm 0.13$ & $2.02 \pm 0.02 \pm 0.17$ & $2.54 \pm 0.03 \pm 0.30$\\
        $\Delta p$ & $1.097 \pm 0.005$ & $1.564 \pm 0.009$ & $2.084 \pm 0.009$ & $2.55 \pm 0.01$\\
        $\Delta p+\Delta r$ & $1.155 \pm 0.005$ & $1.602 \pm 0.009$ & $2.116 \pm 0.009$ & $2.56 \pm 0.01$\\
        {\tt Gaussian} & $1.077 \pm 0.005$ & $1.560 \pm 0.009$ & $2.054 \pm 0.009$ & $2.51 \pm 0.01$\\
        {\tt Argonne} & $1.366 \pm 0.006$ & $1.818 \pm 0.009$ & $2.372 \pm 0.009$ & $2.83 \pm 0.01$\\
        \bottomrule
    \end{tabular*}
\end{table*}

The {\tt Argonne} prediction, although not tuned, remains close to the tuned models, with deviations at the level of $\sim 20$--$25\%$ in the measured $p_T$ window. We interpret this spread as a conservative estimate of the residual model uncertainty associated with the choice of deuteron wavefunction/Wigner kernel in our framework.
Finally, the relative-deviation panels show that the largest discrepancies for all models tend to appear at the highest $p_T$, where the spectra contribute less to the total multiplicity than the low-$p_T$ region, which dominates the integrated yield.
Notably, the residual structure is largely correlated among models, suggesting that remaining differences are driven primarily by the underlying (anti)nucleon production in the generator rather than by the details of the coalescence prescription.
This interpretation is consistent with the fact that our tuned \texttt{PYTHIA} configuration slightly underestimates the \textsf{ALICE} antiproton spectra in the range $p_T \sim 1.5$--$4$~GeV$/c$, which overlaps with the region where we also tend to underestimate the (anti)deuteron spectra.

\subsection{Model predictions for antihelion-3}

In addition to the (anti)deuteron spectra, we also examine the predictions of our \texttt{PYTHIA} tune for the ${}^3\overline{\mathrm{He}}$ spectrum. In our model, a ${}^3\overline{\mathrm{He}}$ nucleus is formed when the coalescence criteria are simultaneously satisfied for all three antinucleon pairs in a $\bar p\bar p\bar n$ configuration.

\begin{figure}
    \centering
    \includegraphics[width=1\linewidth]{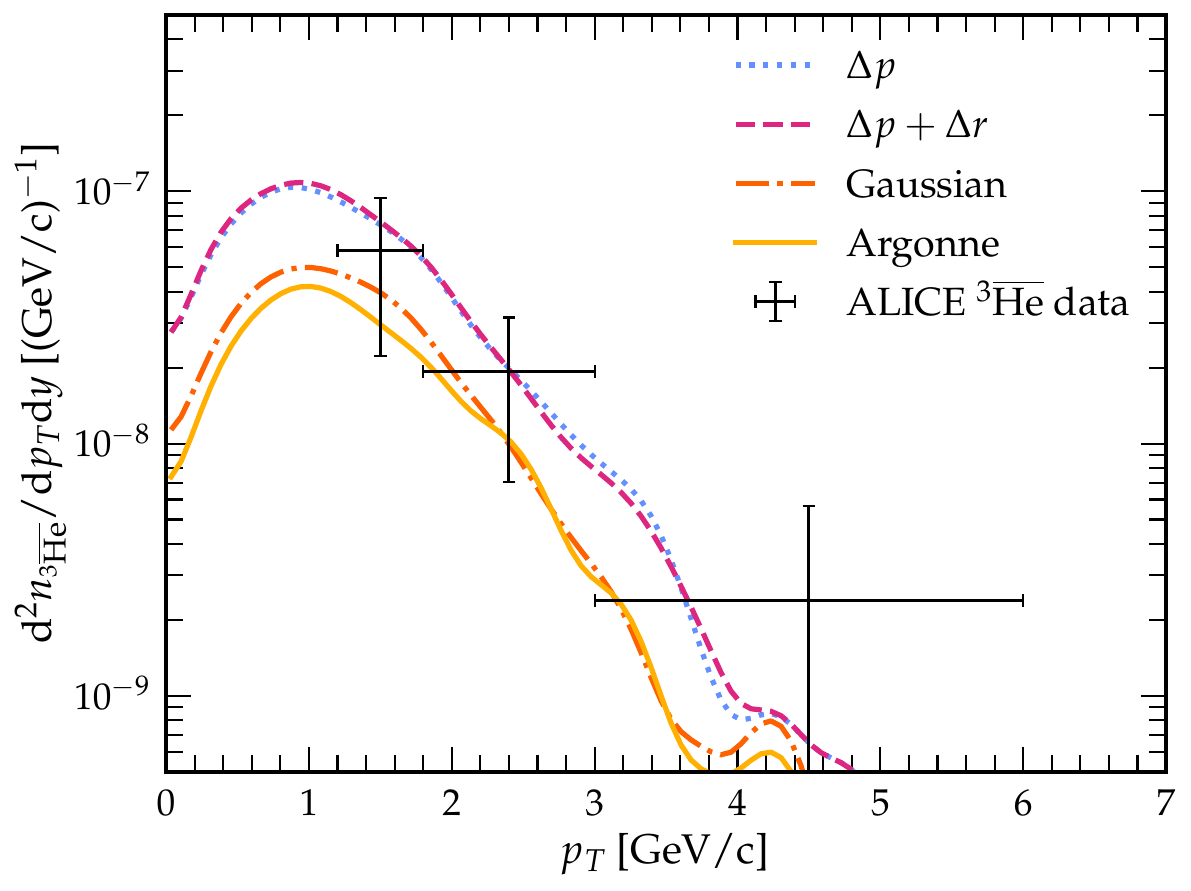}
    \caption{Antihelion-3 $p_T$-differential multiplicity.
    \textsf{ALICE} ${}^3\overline{\rm He}$ spectrum at midrapidity ($|y|<0.5$) for $\sqrt{s}=7$~TeV, compared to the best-fit predictions of the tuned coalescence models ($\Delta p$, $\Delta p+\Delta r$, {\tt Gaussian}) and to the parameter-free {\tt Argonne} implementation.}
    \label{fig:He3barALICE}
\end{figure}

The \textsf{ALICE} Collaboration has released transverse-momentum--differential spectra of ${}^3\overline{\mathrm{He}}$ at midrapidity for $pp$ collisions at $\sqrt{s}=7$~TeV~\cite{ALICE:2025byl} and $13$~TeV~\cite{ALICE:2021mfm}. Figure~\ref{fig:He3barALICE} shows a comparison between the \textsf{ALICE} measurement at $\sqrt{s}=7$~TeV~\cite{ALICE:2025byl} and our predictions for all coalescence models considered, obtained from simulations of $5\times10^9$ events. Within the statistical uncertainties of both the experimental data and the Monte-Carlo simulations, the predictions are consistent with the \textsf{ALICE} results. At large $p_T$ the Monte-Carlo statistics are not sufficient to obtain a smooth theoretical spectrum; however, in this region the experimental uncertainties are also large.

\section{Secondary antideuteron source term}
\label{sec:source_secondary}

Secondary antideuterons are produced in inelastic interactions of primary CRs with the atoms of the ISM. The dominant channels are CR $p$ and He projectiles on H and He targets. The differential source term at Galactic position $\vec x$ (production rate per unit volume, time, and kinetic energy per \emph{n}) is
\begin{widetext}
\begin{equation}
\label{eq:qsec_general}
q_{\overline{\rm D}}^{\rm sec}(K_{\overline{\rm D}},\vec x)
=
\sum_{i,j}
\int_{K_{i,{\rm th}}}^{\infty} dK_i \,
4\pi \, n_{{\rm ISM},j}(\vec x)\,
\phi_i(K_i,\vec x)\,
\frac{d\sigma_{ij\to {\overline{\rm D}}}}{dK_{\overline{\rm D}}}(K_i,K_{\overline{\rm D}}) \,,
\end{equation}
\end{widetext}
where $K_i$ and $K_{\overline{\rm D}}$ denote the kinetic energies per \emph{n} (in GeV/n) of the projectile $i$ and of the produced antideuteron, respectively; $n_{{\rm ISM},j}$ is the number density of the ISM target species $j$; and $\phi_i$ is the interstellar (demodulated) differential CR flux per unit area, time, solid angle, and kinetic energy per \emph{n}.
The sums run over the relevant projectiles and targets; in practice $i\in\{p,{\rm He}\}$ and $j\in\{{\rm H},{\rm He}\}$ dominate, while channels induced by secondary $\bar p$ are subleading.
In a disk-averaged treatment one may adopt constant densities $n_{{\rm ISM,H}}\simeq 1~{\rm cm}^{-3}$ and $n_{{\rm ISM,He}}\simeq 0.1~{\rm cm}^{-3}$, while more refined calculations can use spatially dependent gas maps.

The lower limit $K_{i,{\rm th}}$ in Eq.~\eqref{eq:qsec_general} accounts for the kinematic threshold for antideuteron production implied by baryon-number conservation. For fixed-target $pp$ collisions, antideuteron production requires $\sqrt{s}\gtrsim 6m_p$, which corresponds to a threshold projectile total energy
$E_{p,{\rm th}}\simeq (s-2m_p^2)/(2m_p)\gtrsim 17m_p$ thus a threshold kinetic energy per \emph{n}
$K_{p,{\rm th}}\equiv E_{p,{\rm th}}-m_p \gtrsim 16m_p$.
This energy is substantially higher than for antiproton production and is the main reason why the secondary ${\overline{\rm D}}$ spectrum is strongly suppressed at low $K_{\overline{\rm D}}$, i.e.\ in the energy window most relevant for DM searches. In addition, coalescence selects $\bar p$--$\bar n$ pairs with small relative momentum, so that the production of very energetic antinucleons further suppresses the probability to form a ${\overline{\rm D}}$.

In our framework, the differential production cross section $d\sigma_{ij\to{\overline{\rm D}}}/dK_{\overline{\rm D}}$ is obtained from a Monte-Carlo description of the underlying antinucleon production followed by an event-by-event coalescence prescription. It is convenient to express it as
\begin{equation}
\label{eq:dsig_dK_to_dN_dK_general}
\frac{d\sigma_{ij\to {\overline{\rm D}}}}{dK_{\overline{\rm D}}}(K_i,K_{\overline{\rm D}})
=
\sigma_{ij}^{\rm inel}(K_i)\;
\frac{dN_{\overline{\rm D}}}{dK_{\overline{\rm D}}}(K_i,K_{\overline{\rm D}})\,,
\end{equation}
where $dN_{\overline{\rm D}}/dK_{\overline{\rm D}}$ denotes the ${\overline{\rm D}}$ multiplicity per inelastic $ij$ interaction and $\sigma_{ij}^{\rm inel}$ is the inelastic cross section for the CR projectile $i$ colliding against the ISM target $j$. The quantity $dN_{\overline{\rm D}}/dK_{\overline{\rm D}}$ is obtained from our Monte-Carlo implementation of the coalescence process, evaluated on a grid of 200 incoming proton energies between $\sim 25$ and $10^{7}$~GeV.\footnote{In practice, for the He-induced channels we use the same multiplicity tables as a function of the projectile kinetic energy per \emph{n}, consistently with the superposition approximation adopted for the hadronic interaction.}

For the inelastic cross sections entering Eq.~\eqref{eq:dsig_dK_to_dN_dK_general}, we approximate the proton--nucleus case using an empirical mass-scaling law,
\begin{equation}
\label{eq:sigma_inel_scaling}
\sigma^{\rm inel}_{pX}(K_i)=A_X^{0.8}\,\sigma^{\rm inel}_{pp}(K_i),
\end{equation}
motivated by parameterizations of the form
$\sigma^{\rm inel}_{A_{\rm proj},A_{\rm targ}}\simeq \sigma^{\rm inel}_{pp}\,(A_{\rm proj}A_{\rm targ})^{0.8}$
(see, e.g.,~\cite{Korsmeier:2018gcy,Orusa:2022pvp}).
We use the parameterization of $\sigma^{\rm inel}_{pp}$ reported in Appendix~A of Ref.~\cite{Orusa:2022pvp}, and we rewrite it in Appendix~\ref{appx:XS} for completeness.

Figure~\ref{fig:DbarXS} shows the differential ${\overline{\rm D}}$ production cross section in $pp$ collisions, computed with the {\tt Gaussian} Wigner coalescence model, as a function of the projectile and produced kinetic energies per \emph{n}, $(K_p,\,K_{\overline{\rm D}})$. The cross section vanishes below the kinematic threshold, visible as the sharp cutoff for $K_p\lesssim 20~\mathrm{GeV}/n$, as expected from baryon-number conservation. This kinematic threshold actually almost coincides with the projectile kinetic energy associated to lowest center-of-mass energy allowed by \texttt{PYTHIA}. The implementation of these cross sections in the calculation of the antideuteron flux at Earth is discussed in Ref.~\cite{Stefanuto:2026wxy}. The differential cross-sections were computed for $\bar{p}$ and $\overline{\rm D}$ (for the different coalescence models), and are publicly available on the Zenodo repository~\cite{di_mauro_2026_19099608}.

Although the production cross section increases with the projectile energy, the CR flux decreases approximately as a power law, $\phi_p\propto K_p^{-2.7}$, implying a suppression by about $10^{2.7}$ per decade in energy. Therefore, the convolution in Eq.~\eqref{eq:qsec_general} enhances the relative importance of projectile energies below the multi-TeV region. The resulting secondary source term is shown in Fig.~\ref{fig:Dbarq} for the different collision channels (p--H, p--He, He--H, He--He) and for the different coalescence prescriptions. The source term peaks at $K_{\overline{\rm D}}\sim$ a few GeV/n, and the p--H contribution dominates because of the larger H target density, while the He--He contribution is suppressed (at the level of $\mathcal{O}(30)$) by both the smaller He abundance and the smaller He flux.

Fig.~\ref{fig:Dbarq} also shows that the $\Delta p$ model predicts a larger number of secondary \dbar\ compared to the other coalescence models. In the \textsf{ALICE} analyses, antiprotons originating from weakly decaying hadrons (e.g.\ $\Lambda$ baryons) are removed from the sample. This is typically called feed-down correction. In the simulations used for the source term evaluation, however, such antinucleons must be retained, as they can contribute to total secondary \dbar\ production  in the Galaxy. Antinucleons produced in the decays of weakly decaying hadrons typically originate from displaced vertices and are therefore much more spatially separated from other antinucleons than those produced directly through hadronization or strong resonance decays. In particular, there is basically no chance to make \dbar\ if one antinucleon comes from weakly decaying hadron and the other is generated from another particle decay or is primordial. As a result, coalescence models that include a spatial criterion, such as the $\Delta p+\Delta r$ model, effectively remove this possible contribution. In contrast, the $\Delta p$ model applies only a momentum-space criterion, so this possibility remains, leading to an enhanced, biased secondary \dbar\ yield. Consequently, special care must be taken when applying the $\Delta p$ model to \dbar\ production in astrophysical environments (e.g.\ DM annihilation or secondary production in cosmic-ray interactions), as it may lead to the formation of a non-negligible fraction of non-physical \dbar.

\begin{figure}
    \centering
    \includegraphics[width=\linewidth]{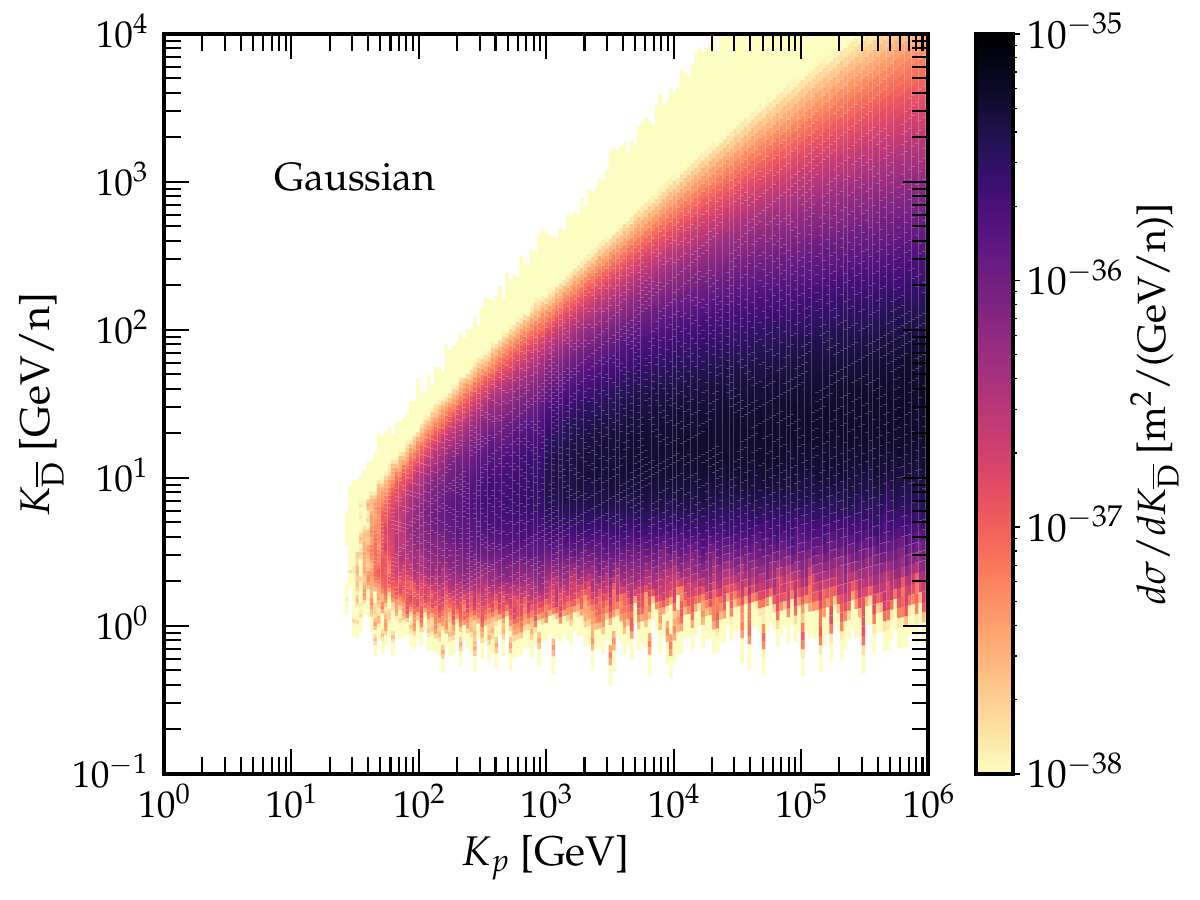}
    \caption{Differential antideuteron production cross section $d\sigma/dK_{\overline{\rm D}}$ in $pp$ collisions
    (color scale) in the laboratory frame as a function of the projectile and produced kinetic energies per nucleon $(K_p,\,K_{\overline{\rm D}})$. The result is computed with the {\tt Gaussian} Wigner coalescence model.}
    \label{fig:DbarXS}
\end{figure}

\begin{figure}
    \centering
    \includegraphics[width=\linewidth]{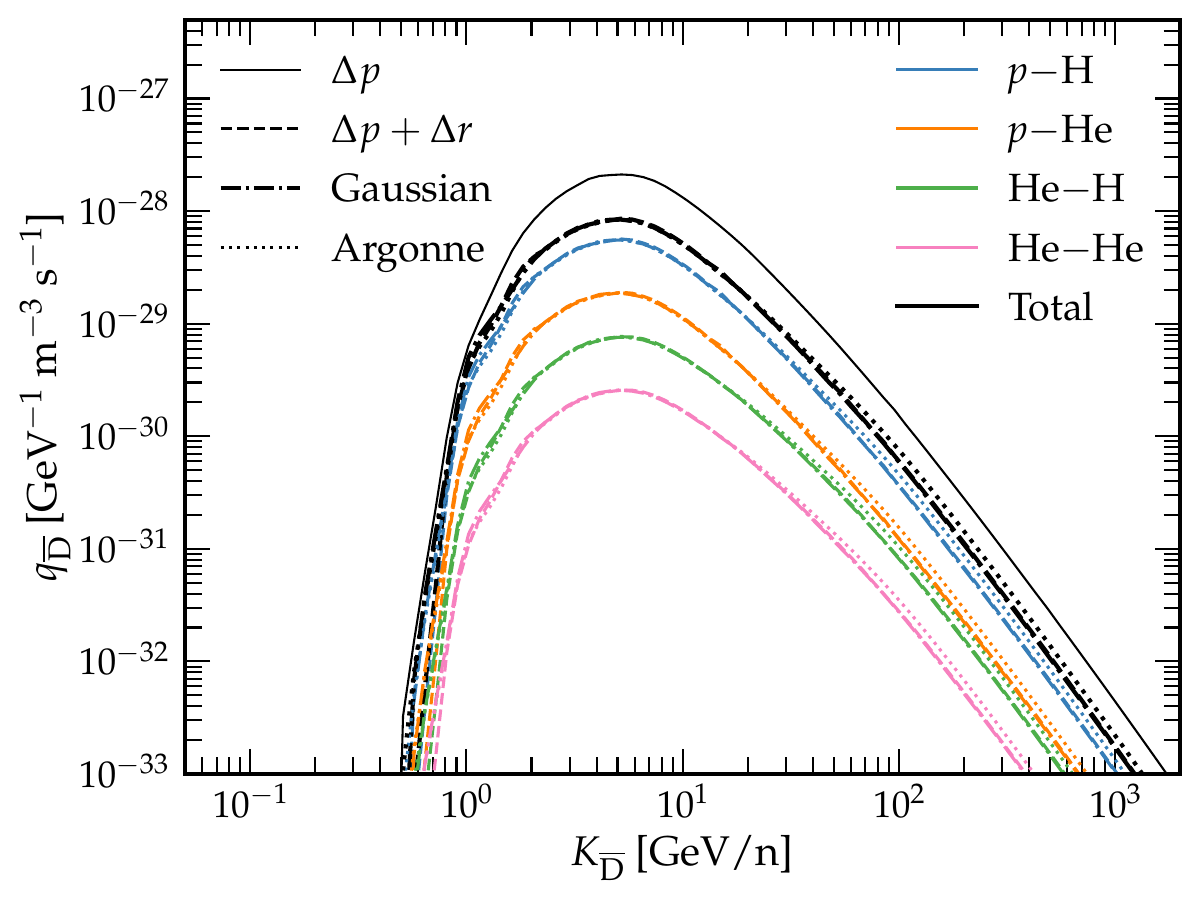}
    \caption{Secondary antideuteron source term
    $q_{\overline{\rm D}}(K_{\overline{\rm D}})$ for secondary ${\overline{\rm D}}$ production, split by collision channel ($p$--H, $p$--He, He--H, He--He) and total (black). For each channel, different line styles correspond to different coalescence prescriptions: $\Delta p+\Delta r$ (dashed), {\tt Gaussian} (dot-dashed), {\tt Argonne} (dotted). For the $\Delta p$ model, we show only the total source term (solid).}
    \label{fig:Dbarq}
\end{figure}

\section{Discussion and conclusions}
\label{sec:conclusions}

In this work we studied the production of light antinuclei in hadronic collisions with the goal of reducing the dominant source-term uncertainty in CR antideuteron (and, prospectively, antihelium) searches for DM. The key difficulty is that antinucleus formation is intrinsically non-perturbative: it depends both on the underlying antinucleon production in QCD hadronization and on the subsequent coalescence of $\bar p$ and $\bar n$ into a bound state. Our strategy was therefore to (i) calibrate the antinucleon production with collider data over a wide energy range and (ii) test a set of increasingly realistic coalescence prescriptions, anchoring their parameters to accelerator measurements in different production environments.

On the generator side, we employed \texttt{PYTHIA}~8.315 as baseline and implemented antinucleus formation directly on the event record. Since coalescence is sensitive to antinucleon correlations and, for models with spatial information, to production vertices, it is crucial that the generator reproduces the relevant single-particle observables. For this reason, the tune was validated against a compilation of antiproton multiplicities across fixed-target and collider energies, as well as against differential $p_T$ multiplicities (and invariant cross sections) from \textsf{NA61}, \textsf{NA49} and \textsf{ALICE}. The outcome is a generator setup that reproduces both the global antiproton multiplicity and spectra, and the key kinematic shapes over more than three orders of magnitude in the equivalent beam energy, which is essential for reliable predictions of low-$K$ antinuclei yields relevant for DM searches. In the absence of dedicated \textsf{LHC} antineutron measurements, we adopted isospin symmetry for $\bar p$ and $\bar n$ production at high energy, consistent with inclusive hadron data and with the default \texttt{PYTHIA} treatment.

We then compared several coalescence implementations that span the range of common approaches in the literature. First, we considered the standard event-by-event ``phase-space cutoff'' prescriptions: the $\Delta p$ model, where a $\bar p$--$\bar n$ pair forms an antideuteron if its relative momentum in the pair center-of-mass frame satisfies $|\Delta \vec p|<p_{\rm coal}$, and the $\Delta p+\Delta r$ extension, where one additionally requires a spatial proximity cut $|\Delta \vec r|<r_{\rm coal}$. These models retain dynamical correlations from the generator and, in the $\Delta r$ case, suppress coalescence between promptly produced antinucleons and those originating from displaced vertices (e.g.\ weak decays), which are naturally separated in space--time. Second, we implemented quantum-mechanical coalescence in the Wigner formalism, where the formation probability is obtained by projecting the two-particle phase-space distribution onto the bound-state Wigner function. We studied both a {\tt Gaussian} benchmark (controlled by an effective momentum/size scale $\delta$) and a more predictive {\tt Argonne} implementation based on the Argonne $v_{18}$ deuteron wavefunction, constrained by nucleon--nucleon scattering data and the deuteron binding energy.

The core calibration step of this paper is a simultaneous fit to the \textsf{ALICE} $p_T$-differential multiplicities of deuterons and antideuterons measured at midrapidity ($|y|<0.5$) at $\sqrt{s}=0.9$, $2.76$, $7$ and $13$~TeV. We performed global $\chi^2$ fits on a grid of the relevant coalescence parameter for each model, comparing the Monte-Carlo predictions in the same kinematic bins as the data. To avoid parameter degeneracies that are known to arise when both a source-size parameter and a bound-state parameter are left free (especially in Gaussian/Wigner setups), we fixed the spatial scale to a benchmark $3~\mathrm{fm}$, comparable to the deuteron size inferred from its charge radius, and fitted only a single parameter per model: $p_{\rm coal}$ for $\Delta p$ and $\Delta p+\Delta r$ (with $r_{\rm coal}=3~\mathrm{fm}$), and $\delta$ for the Gaussian/Wigner setup (with $\sigma=3~\mathrm{fm}$). In addition to the \textsf{ALICE} calibration, we compared the extracted parameters to those obtained in Ref.~\cite{DiMauro:2024kml} from the \textsf{ALEPH} measurement of the $\overline{\mathrm D}$ multiplicity in hadronic $Z$ decays at $\sqrt{s}=m_Z$, an environment often used as a proxy for DM annihilation into quarks.

A central result is that the coalescence scales extracted from \textsf{ALICE} $pp$ data are consistent with those inferred from \textsf{ALEPH} within uncertainties, despite the very different underlying environments and center-of-mass energies. Quantitatively, we find $p_{\rm coal}\simeq 0.20~\mathrm{GeV}$ for the $\Delta p$ model and a slightly larger value, $p_{\rm coal}\simeq 0.21~\mathrm{GeV}$, when a spatial cut is imposed, as expected because the $\Delta r$ requirement removes a fraction of pairs that would otherwise coalesce. In the {\tt Gaussian} setup, the best-fit scale is $\delta\simeq 1.7~\mathrm{fm}$, close to the value obtained from \textsf{ALEPH} within the larger uncertainties of the \textsf{LEP} measurement. The \textsf{ALICE}-based determination achieves sub-percent statistical precision in the fitted parameter (driven by the small experimental errors and large Monte-Carlo samples), while the \textsf{ALEPH} calibration remains at the $\mathcal{O}(10\%)$ level because of the larger uncertainty on the measured multiplicity. Importantly, once tuned, the $\Delta p$, $\Delta p+\Delta r$ and {\tt Gaussian} prescriptions lead to very similar antideuteron spectra, typically differing by $\lesssim 10\%$ over the measured $p_T$ range. The {\tt Argonne} prediction, although parameter free and not tuned to \textsf{ALICE}, remains compatible at the $\sim 20$--$25\%$ level. We interpret this spread as a conservative estimate of the residual model dependence associated with the choice of bound-state wavefunction/Wigner kernel, once the antinucleon production is properly calibrated.

We also find that the global fit quality improves significantly when excluding the $\sqrt{s}=13$~TeV dataset, while the best-fit parameters shift only mildly. This pattern indicates that the coalescence scales are stable and that the remaining tension is more plausibly linked to the generator-level description of (anti)nucleon kinematics at the highest energy (notably at higher $p_T$) than to the coalescence prescription itself. The correlated residual structure across coalescence models supports this interpretation and motivates future work combining (i) further refinements of the high-energy baryon-production tune and (ii) additional differential constraints (e.g.\ multiplicity- or event-activity-dependent light-nuclei measurements) to stress-test the universality of the coalescence description beyond the inclusive midrapidity spectra used here.

Ref.~\cite{Kachelriess:2020uoh} likewise found that a conventional coalescence model implemented within the Wigner formalism can describe the \textsf{ALICE} and \textsf{ALEPH} data for \(\overline{\rm D}\). Nevertheless, that analysis did not make use of the full event-by-event spatial and momentum information of the antinucleon pairs produced by the event generator. Instead, it employed an integrated version of the Wigner formalism, see Eq.~\eqref{eq:wignerDspectrum_rephr}, based on the Monte Carlo antinucleon momentum distributions and an effective treatment of the source size, thereby averaging over the detailed spatial structure of the \(\bar p\)-\(\bar n\) emission process.

Finally, we outlined how the calibrated event-by-event framework connects to the secondary antideuteron source term in the Galaxy: the production rate is obtained by convolving the interstellar CR fluxes with ISM gas densities and the differential production cross sections, which in our Monte-Carlo approach can be written as $d\sigma_{ij\to{\overline{\rm D}}}/dK_{\overline{\rm D}}=\sigma_{ij}^{\rm inel}\,dN_{\overline{\rm D}}/dK_{\overline{\rm D}}$. The collider-calibrated coalescence parameters derived in this work therefore provide a controlled and transferable input for the antideuteron production multiplicities entering astrophysical calculations. We however stress that the $\Delta p$ model would generate a non-negligible fraction of non-physical secondary \dbar.

In conclusion, by jointly confronting coalescence models with \textsf{ALICE} $pp$ spectra and \textsf{ALEPH} hadronic-$Z$ data, we provide quantitative evidence that a common coalescence description can be applied across widely different energies and production mechanisms. This substantially strengthens the physical basis of antinucleus source-term predictions and improves the robustness of interpreting forthcoming searches with \textsf{AMS-02} and \textsf{GAPS}, where even a handful of low-$K$ antideuteron candidates could carry decisive information about particle DM.

\begin{acknowledgments} 
M.D.M., F.D., N.F., J.K. and L.S.~acknowledge support from the research grant {\sl TAsP (Theoretical Astroparticle Physics)} funded by Istituto Nazionale di Fisica Nucleare (INFN). M.D.M.~and J.K.~acknowledge support  from the Italian Ministry of University and Research (MUR), PRIN 2022 ``EXSKALIBUR – Euclid-Cross-SKA: Likelihood Inference Building for Universe’s Research'', Grant No. 20222BBYB9, CUP I53D23000610 0006, and from the European Union -- Next Generation EU. J.K.~acknowledges support from the Italian Space Agency through the ASI INFN agreement n. 2018-28-HH.0: “Partecipazione italiana al GAPS - General AntiParticle Spectrometer”. N.F.~is supported by the Italian Ministry of University and Research (\textsc{mur}) via the PRIN 2022 Project No. 20228WHTYC – CUP: D53C24003550006 and from the European Union -- Next Generation EU. F.B. acknowledges support from the European Research Council under the European Union’s Horizon 2020 research and innovation programme through the ERC-H2020-STG CosmicAntiNuclei project (GA n. 950692).

\end{acknowledgments}

\bibliographystyle{apsrev4-1}
\bibliography{paper.bib}

\newpage

\onecolumngrid
\appendix
\section{Additional PYTHIA tuning details}
\label{appx:tuning}

\begin{figure}[t]
    \centering
    \includegraphics[width=0.45\linewidth]{plots/tuning/spec_NA61_31GeV_pbar.pdf}
    \includegraphics[width=0.45\linewidth]{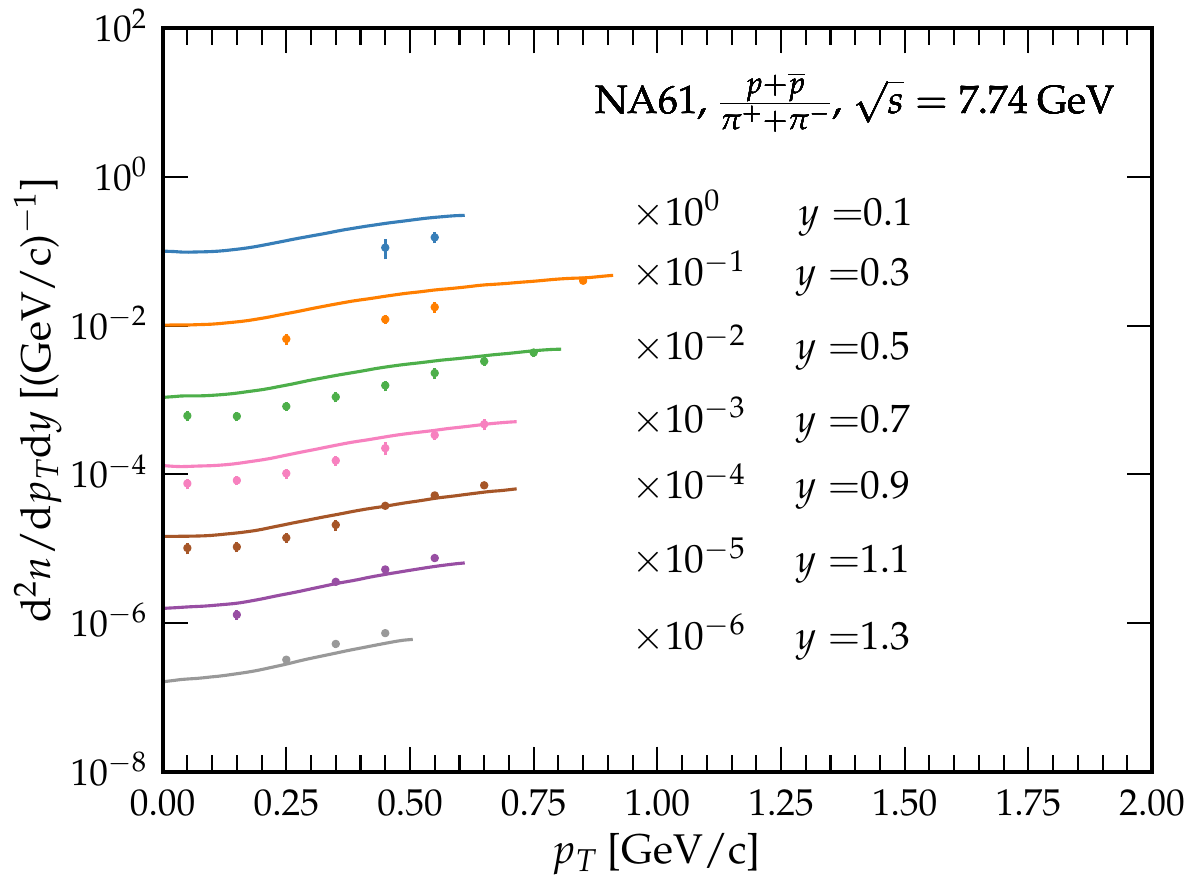}
    \includegraphics[width=0.45\linewidth]{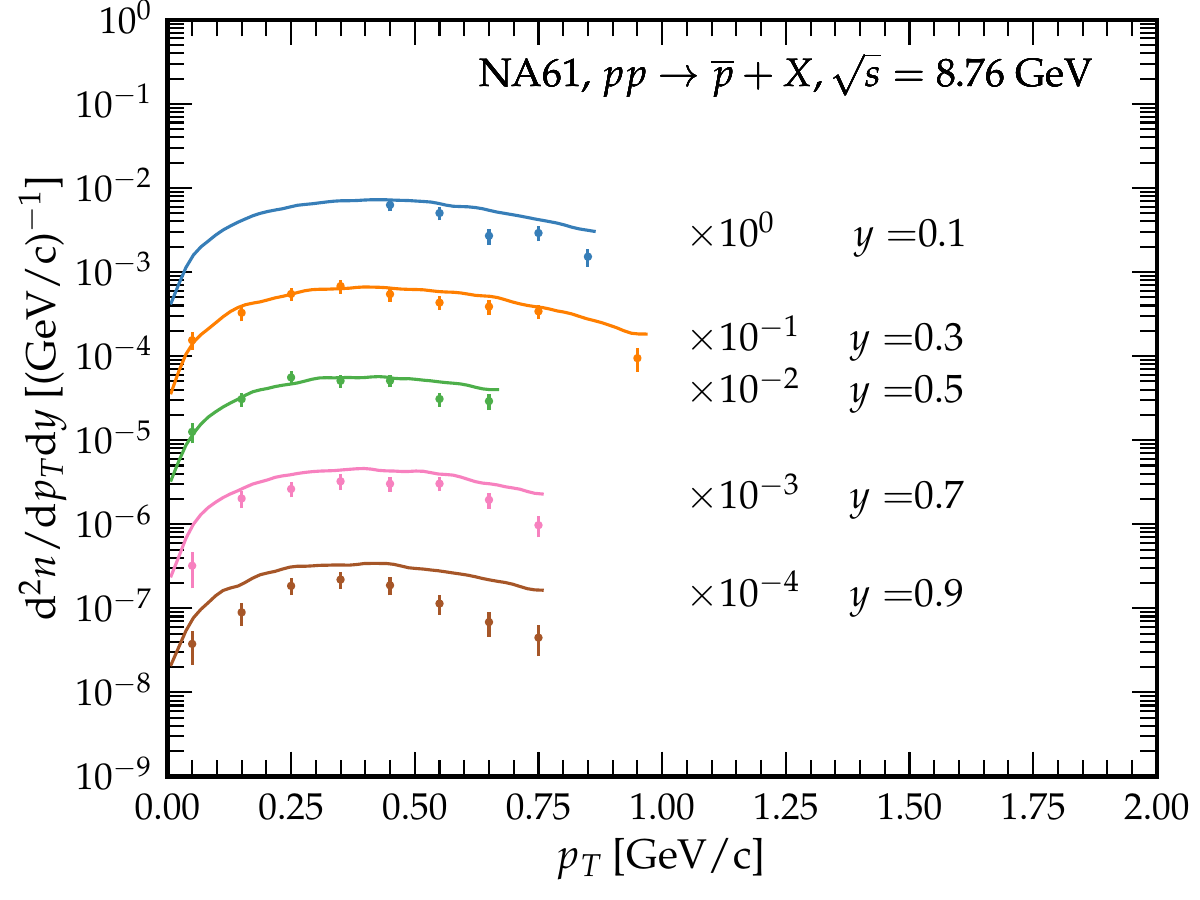}
    \includegraphics[width=0.45\linewidth]{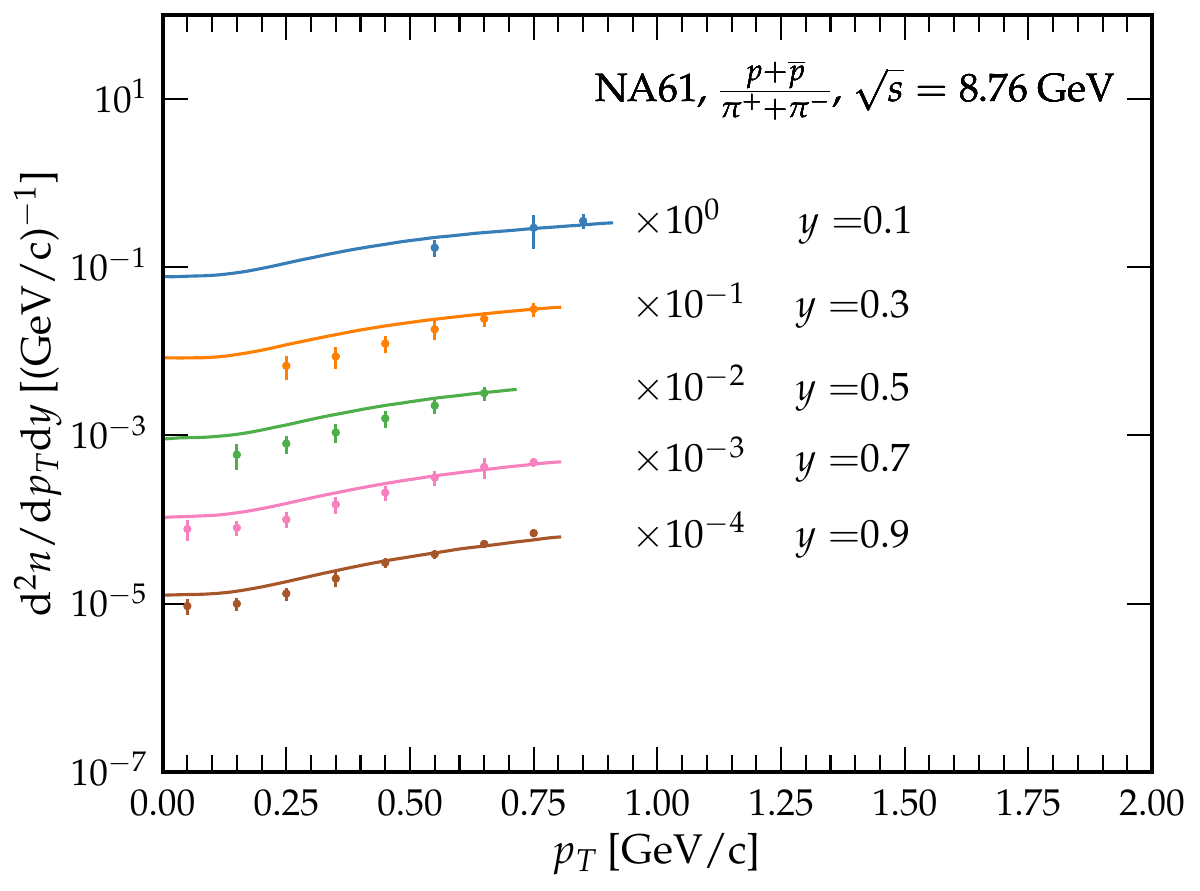}
    \includegraphics[width=0.45\linewidth]{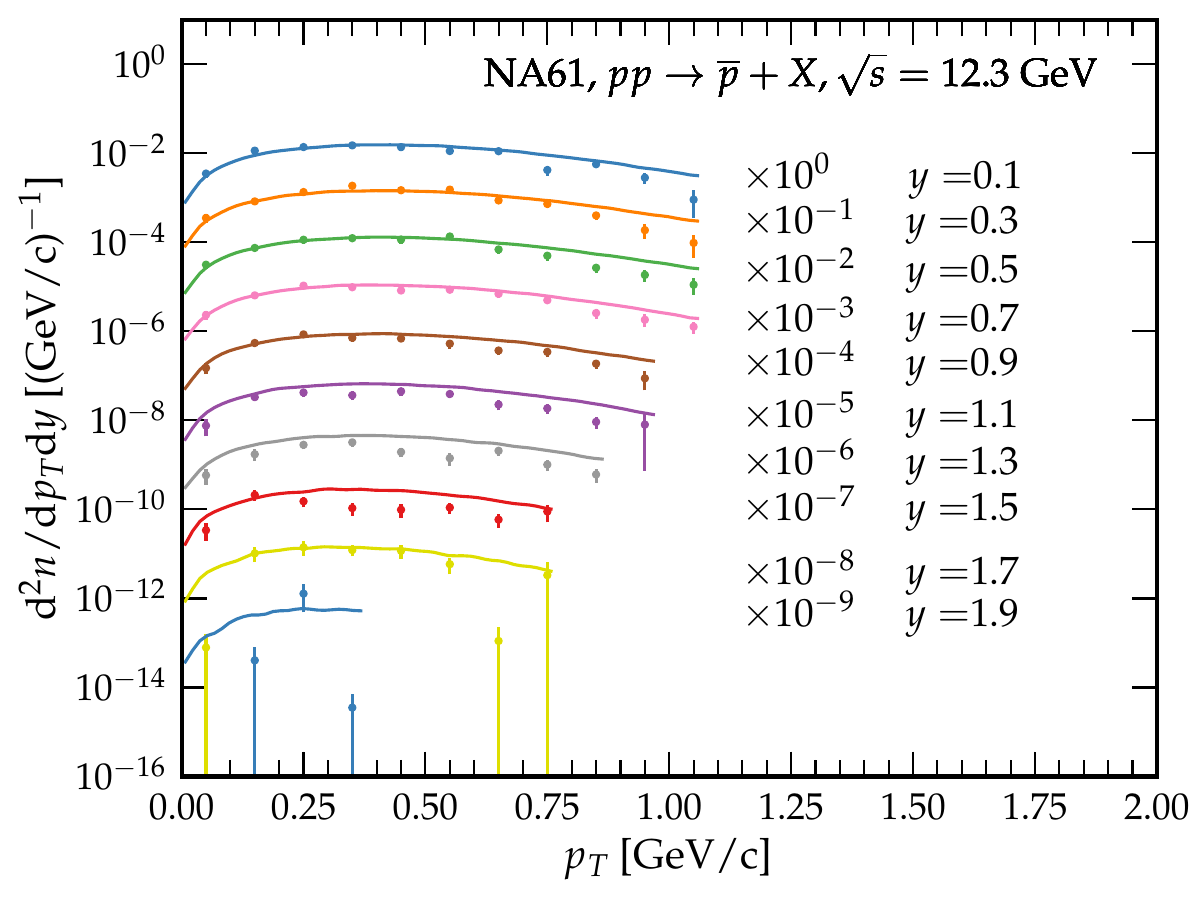}
    \includegraphics[width=0.45\linewidth]{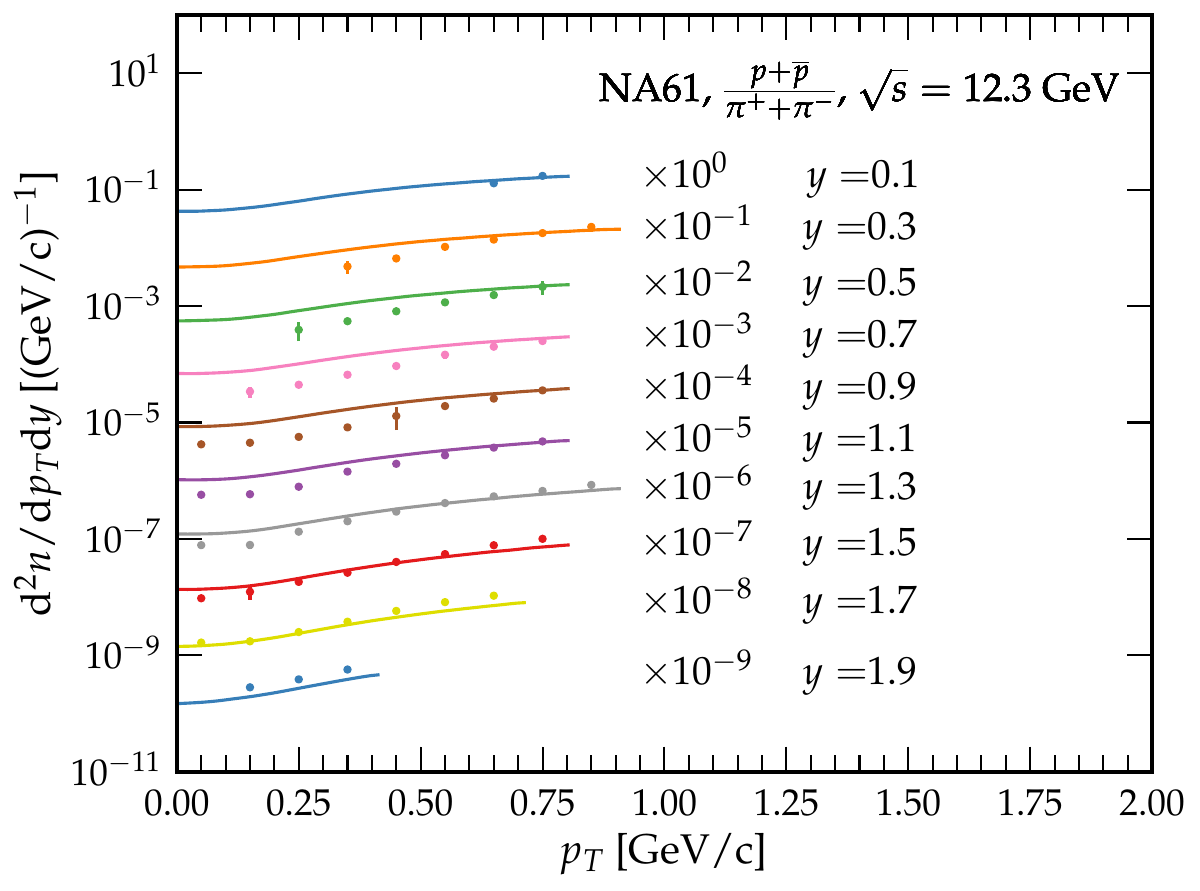}
    \includegraphics[width=0.45\linewidth]{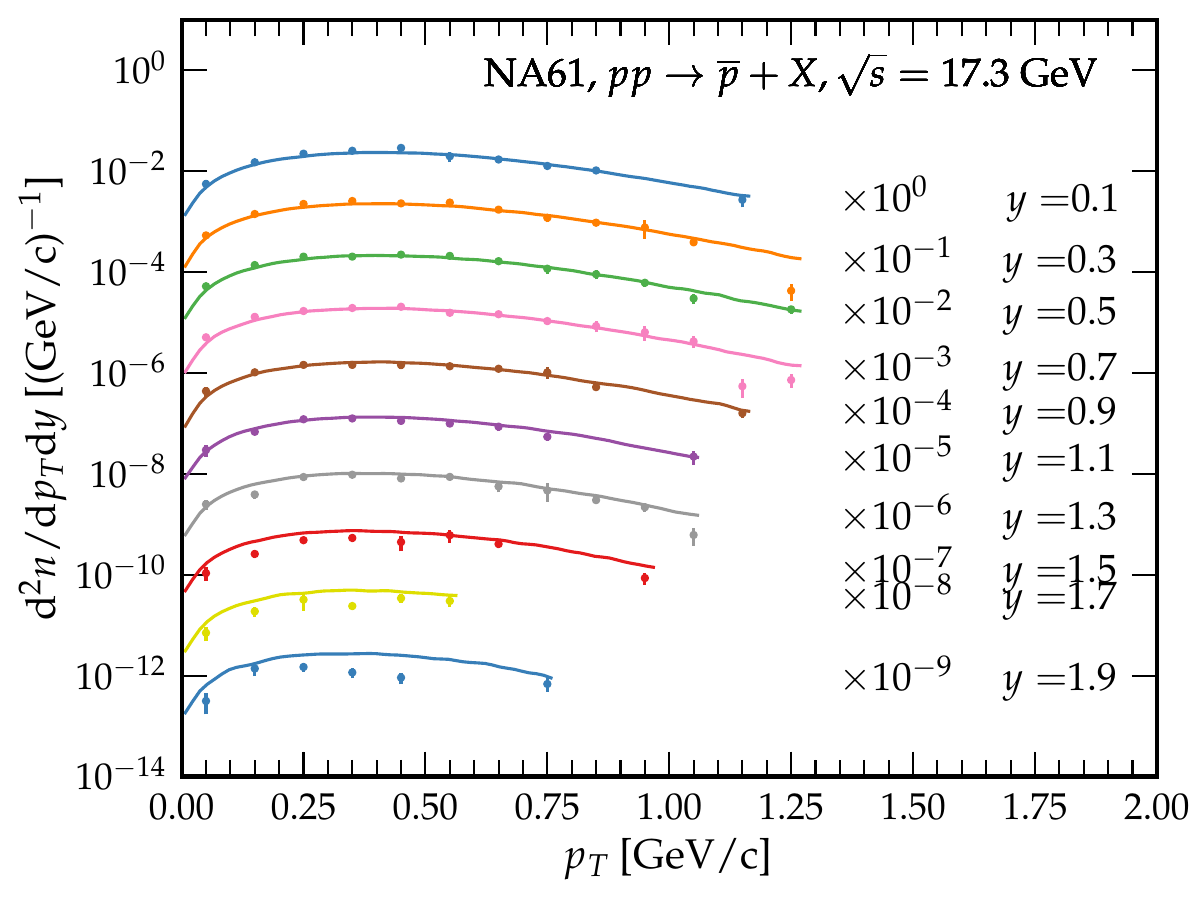}
    \includegraphics[width=0.45\linewidth]{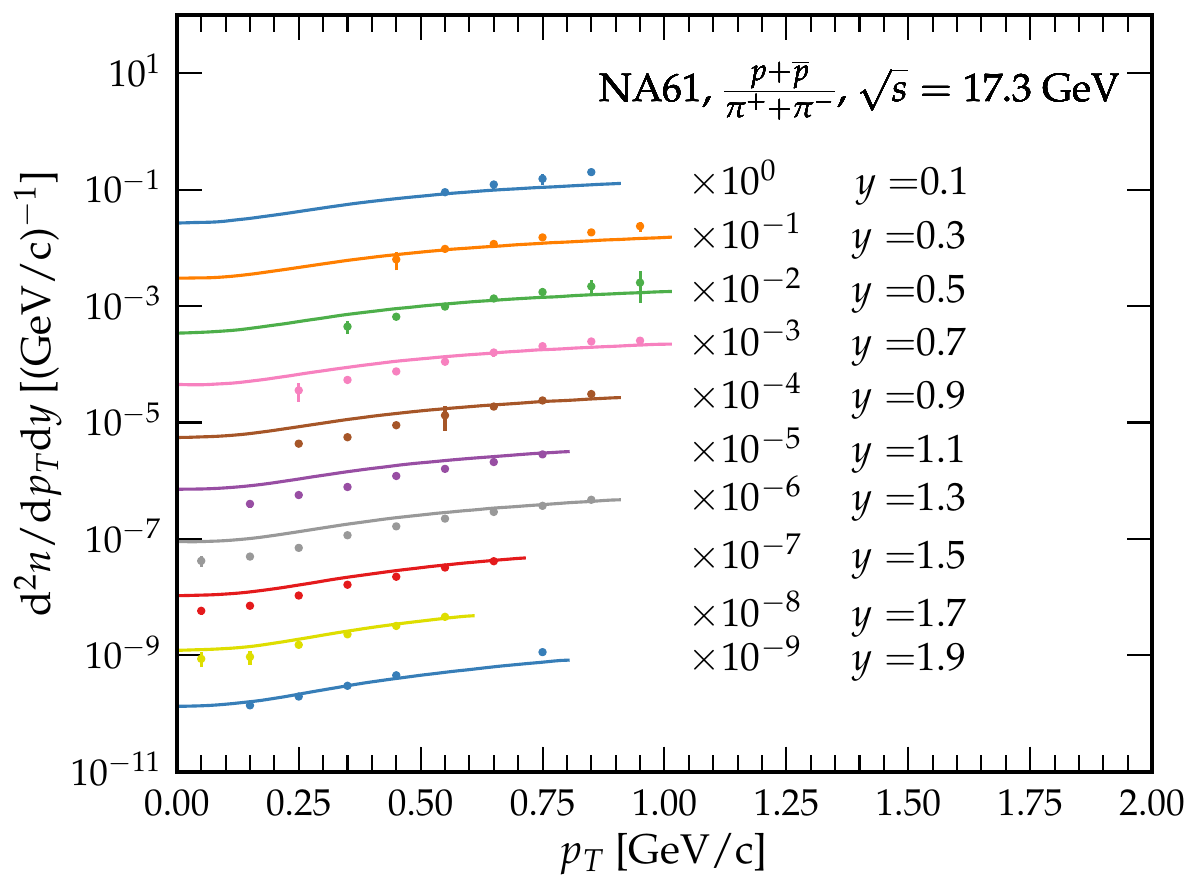}
    \caption{Transverse momentum spectra of antiprotons (left panels) and proton-over-pion ratios (right panels) generated with our tuning of \texttt{PYTHIA}, compared to \textsf{NA61} data~\cite{NA61SHINE:2017fne} at different $y$ in fixed-target $pp$ collisions at $\sqrt{s} = 7.74$ GeV (top panels), $8.76$ GeV (middle top panels), $\sqrt{s} = 12.3$ GeV (middle bottom panels) and $17.3$ GeV (bottom panels).}
    \label{fig:NA61pbar}
\end{figure}

\begin{figure}[t]
    \centering
    \includegraphics[width=0.45\linewidth]{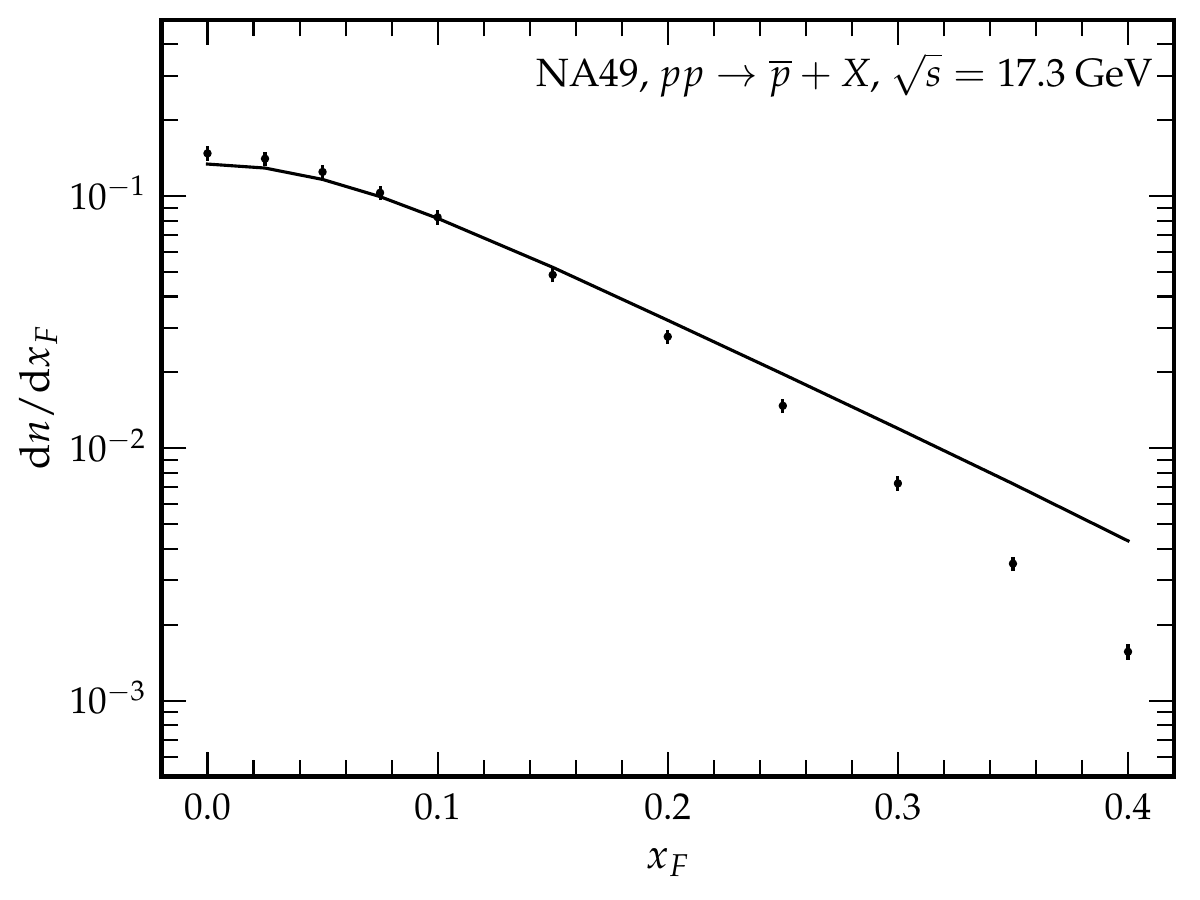}
    \caption{$p_T$-integrated $x_F$-spectrum $dn/dx_F$ of antiprotons generated with our tuning of \texttt{PYTHIA}, compared to \textsf{NA49} data~\cite{NA49:2009brx}, in fixed-target $pp$ collisions at $\sqrt{s}=17.3$ GeV.}
    \label{fig:NA49xFpbar}
\end{figure}

In the left panels of Fig.~\ref{fig:NA61pbar} we compare the predicted antiproton transverse momentum $p_T$ spectra for a range of rapidities $y$ with the measurements carried by \textsf{NA61}, while in the right panels we do the same for the $p_T$-spectra of the ratio between protons and pions $(p+\overline{p})/(\pi^++\pi^-)$. The panels show, from top to bottom, results at $\sqrt{s} = 7.74$ GeV ($E_p^{\rm{LAB}}=31$ GeV), $\sqrt{s}=8.76$ GeV ($E_p^{\rm{LAB}}=40$ GeV), $\sqrt{s}=12.3$ GeV ($E_p^{\rm{LAB}}=80$ GeV) and $\sqrt{s}=17.3$ GeV ($E_p^{\rm{LAB}}=158$ GeV). In Fig.~\ref{fig:NA49xFpbar}, we compare the predicted $p_T$-integrated antiproton $x_F$-spectrum $dn/dx_F$ with the measurements reported by \textsf{NA49} at $\sqrt{s}=17.3$ GeV ($E_p^{\rm LAB}=158$ GeV).

These plots confirms what was already stated in Sec.~\ref{sec:tuning}, which is that our \verb|PYTHIA| tuning predicts well the measured properties of antiprotons produced in $pp$ collisions across a broad range of input proton energies $E_p^{\rm{LAB}}$.

\section{Additional cross sections details}
\label{appx:XS}

In this Appendix, we report the parametrization of the total inelastic cross-section for $pp$ collisions, as calculated in Ref.~\cite{Orusa:2022pvp}. This is done by first parameterizing the total collision ($\sigma_{\rm tot}^{pp}$) and elastic ($\sigma_{\rm el}^{pp}$) cross-sections where the following function is used, which depends on the Mandelstam variable $s$:
\begin{equation}
    \label{eq:ppXS}
    \sigma_{\rm tot/el}^{pp}(s) = Z^{pp}+B^{pp}\log^2(s/s_M) + Y_1^{pp}(s_M/s)^{\eta_1}-Y_2^{pp}(s_M/s)^{\eta_2}\;,
\end{equation}
where $B^{pp}=\pi(\hbar c)^2/M^2$, $s_M=(2m_p+M)^2$ and the other parameters are summarized in Tab.~\ref{tab:XSparams}. The resulting inelastic cross-section is then simply $\sigma_{\rm inel}^{pp}(s) = \sigma_{\rm tot}^{pp}(s)-\sigma_{\rm el}^{pp}(s)$.

\begin{table}[h]
    \centering
    \begin{tabular}{|c|c|c|}
        \hline
        Parameter & Total & Elastic \\
        \hline
        $M$ & 1.589 & 3.094 \\
        \hline
        $Z^{pp}$ & 59.58 & 21.34 \\
        \hline
        $Y_1^{pp}$ & 0.890 & 2.667 \\
        \hline
        $Y_2^{pp}$ & 19.35 & 14.21 \\
        \hline
        $\eta_1$ & 2.543 & 1.003 \\
        \hline
        $\eta_2$ & $-0.0895$ & $-0.0327$ \\
        \hline
    \end{tabular}
    \caption{Parameters for the total and elastic proton scattering cross sections according to Eq.~\eqref{eq:ppXS}.}
    \label{tab:XSparams}
\end{table}

\end{document}